\documentclass{aa}
\usepackage{natbib}         % For citations
\usepackage{graphicx,color}
\usepackage{txfonts}

\newcounter{ref}
\newcommand{\p}[1]{{\color{black} {#1}}\color{black}{}}

\newcommand{\rmd}{ {\mathrm d} }

\newcommand{\uvec}[1]{ \hat{\bf #1} }

\newcommand{\dVxdt}{ (\rmd V_R / \rmd t)_{\rm fit}}

\newcommand{\betap}{\beta_{\rm p}}

\newcommand{\insitu}{\textit{in situ}}
\newcommand{\Np}{N_{\rm p}}

\newcommand{\tin}{t_{\rm in}}
\newcommand{\tout}{t_{\rm out}}
\newcommand{\Vc}{V_{\rm c}}
\newcommand{\nMCp}{20}

\begin{document}

\authorrunning{Gulisano \textit{et al.}}

\title{Expansion of magnetic clouds in the outer heliosphere}
\subtitle{}

\author{A.M. Gulisano \inst{1,2,3}, P. D\'emoulin \inst{4},
S. Dasso \inst{2,1}, L. Rodriguez \inst{5}} \offprints{A.M. Gulisano} \institute{ $^{1}$
Instituto de Astronom\'\i a y F\'\i sica del Espacio, CONICET-UBA,
CC. 67, Suc. 28, 1428 Buenos Aires, Argentina \email{agulisano@iafe.uba.ar, sdasso@iafe.uba.ar}\\
$^{2}$ Departamento de F\'\i sica, Facultad de Ciencias Exactas y
Naturales, Universidad de Buenos Aires, 1428 Buenos Aires, Argentina \email{dasso@df.uba.ar}\\
$^{3}$ Instituto Ant\'artico Argentino (DNA), Cerrito 1248,CABA,
       Argentina\\
$^{4}$ Observatoire de Paris, LESIA, UMR 8109 (CNRS),
       F-92195 Meudon Principal Cedex, France \email{Pascal.Demoulin@obspm.fr}\\
$^{4}$ \p{Solar-Terrestrial Center of
 Excellence - SIDC, Royal Observatory of Belgium, Brussels, Belgium} \email{rodriguez@oma.be}\\
}
   \date{Received ***; accepted ***}

% \abstract{}{}{}{}{}    % 5 {} token are mandatory
   \abstract
   {A large amount of magnetized plasma {is} frequently ejected
   from the Sun as {coronal mass ejections} (CMEs).  {Some} of
   these ejections are detected in the solar wind as
   magnetic clouds (MCs) {that} have flux rope signatures.}
   {Magnetic {clouds are structures that typically} expand in the inner heliosphere.  {We} derive the expansion properties of MCs
   in the outer heliosphere from {one to five astronomical units} to compare them
   {with those} in the inner heliosphere.}
   {We analyze MCs observed by the Ulysses spacecraft
   using \insitu\ magnetic field and plasma measurements.  The
   MC boundaries are defined in the MC frame after defining the
   MC axis with a minimum variance method applied only to the flux
   rope structure.  As in the inner heliosphere, a large fraction of
   the velocity profile within MCs is close to a linear function of time.  This
   {is indicative of} a self-similar expansion and a MC size that locally follows
   a power-law of the solar distance with an exponent called $\zeta$.
   We derive the value of $\zeta$ from the \insitu\ velocity data.}
   {We analyze separately the non-perturbed
   MCs (cases {showing} a linear velocity profile almost for the full event), and
   perturbed MCs (cases {showing} a strongly distorted velocity profile).
   We find that non-perturbed MCs expand with a similar
   non-dimensional expansion rate ($\zeta=1.05\pm 0.34$), \p{i.e.}
   slightly faster than {at} the solar distance and in the inner
   heliosphere ($\zeta = 0.91\pm 0.23$).  The subset of
   perturbed MCs expands, as in the inner heliosphere, {at} a
   {significantly}
   lower rate and with a larger dispersion ($\zeta=0.28\pm 0.52$) as expected
   from the temporal evolution found in numerical simulations.  This local measure
   of the expansion  also {agrees} with the distribution with distance of MC size,
   mean magnetic field, and plasma parameters.  The MCs
  {interacting} with a strong field region, e.g. another MC, have the
   most variable expansion rate (ranging from compression to
   over-expansion). }
{}

        \keywords{Sun: magnetic fields, Magnetohydrodynamics (MHD), Sun: coronal mass ejections (CMEs), Sun: solar wind, Interplanetary medium }

\maketitle

%%%%%%%%%%%%%%%%%%%%%%%%%%%%%%%%%%%%%%%%%%%%%%%%%%%%%%%%%%%%%%%%%%%%%%%%%%%%%%%%%%%%%
\section{Introduction} %%%%%%%%%%%%%%%%%%%%%%%%%%%%%
\label{Introduction}

% {\S\bf What are MCs}\\
Magnetic clouds (MCs) are a subset of interplanetary coronal mass
ejections, {that show} to an observer at rest in the heliosphere:
the large and coherent rotation of the magnetic field vector, a
magnetic field intensity {higher} than in their surroundings, and a
low proton temperature \p{\cite[e.g.,][]{Klein82,Burlaga95}.} They
transport huge helical magnetic structures {into} the heliosphere,
and are believed to be the most efficient mechanism {for releasing}
magnetic helicity from the Sun \p{\cite[][]{Rust94,Low97}.} They are
also the most geoeffective events {to erupt} from the Sun, and {are}
associated with observations of strong flux decreases {for} galactic
cosmic rays
\p{\cite[][]{Tsurutani1988,Gosling1991,Webb2000,Cane2000,Cyr2000,Cane2003}.}

% {\S\bf Expansion}\\
Despite the increasing interest {in improving our} knowledge of MCs
{in} the {past few} decades, {many} properties {remain
unconstrained}, as {such} their dynamical evolution {as} they travel
in the solar wind (SW). In particular, during their evolution, MCs
{undergo} a significant expansion
\p{\cite[e.g.,][]{Klein82,Farrugia93}.} Observations of MCs at
different heliodistances show that their radial size increases with
the distance to the Sun \cite[e.g.,][]{Kumar96,Bothmer98,Leitner07}.
More generally, interplanetary coronal mass ejections (ICMEs) are
also known to {undergo} a global expansion in {the} radial direction
with heliodistance \cite[e.g.,][]{Wang05b,Liu05}. The expansion
properties of MCs and ICMEs then differ significantly from the
stationary Parker's solar wind.

% {\S\bf $V_{R,\rm out}-V_{R,\rm in}$: good proxy ? }\\
From in situ observations of MCs, the plasma velocity profile
typically {decreases as} the MC passes through the spacecraft,
showing a faster speed at the beginning (when the spacecraft is
going into the cloud, $V_{R,\rm in}$) and a slower speed at the
 end \p{\cite[when the spacecraft is going out of the cloud, $V_{R,\rm out}$, e.g.,][]{Lepping03,Gulisano10}.} Moreover,
  the velocity profile inside MCs typically shows a linear profile, which is expected for a self-similar
  expansion \p{\citep{Farrugia93,Shimazu02,Demoulin08,Demoulin09}.}
The time variation of the size of the full MC can be approximated with
$\Delta \p{V_{MC}}=V_{R,\rm out}-V_{R,\rm in}$. However,
%\p{this quantity is not describing the intrinsic expansion of the fluid, since}
\p{a parcel of fluid a factor $f$ times smaller than the MC would
 typically have a difference in velocity $\Delta V=f \; \Delta
V_{MC}$ (because of the linear profile).}

Then, as $\Delta V$ is size dependent, it is not {the} intrinsic
expansion rate of the parcels of fluid.  Moreover, the regions near
the MC boundaries typically {undergo} the strongest perturbations,
so that $\Delta V$ is not a reliable measurement of the flux-rope
expansion rate.

% {\S\bf Why we introduce $\zeta$} \\
{To measure with greater accuracy} the expansion of plasma in MCs,
\cite{Demoulin08} introduced a non-dimensional expansion rate
(called $\zeta$, see Sect.~\ref{E-zeta}), which is computed from the
local insitu measurements.  It was shown that, if $\zeta$ is
constant in time, then $\zeta$ also {determines} the exponent of the
self-similar expansion, i.e. the size of a parcel of fluid as $r(D)
\sim r_0 D^\zeta$, with $D$ the distance to the Sun.

% {\S\bf Summary of previous results with $\zeta$ at $D \leq 1$~AU}\\
From the analysis of the non-dimensional expansion rate ($\zeta$)
for a large sample of MCs observed in the inner heliosphere by the
spacecrafts Helios 1 and 2, \cite{Gulisano10} found that there are
two populations of MCs, one {that has} a linear profile of plasma
velocity {throughout} almost the {entire} passage of the cloud {by}
the spacecraft (the non-perturbed subset) and another population
{that has} a strongly distorted velocity profile (the perturbed
subset). The presence of the second population was interpreted as a
consequence of the interaction of a fraction of the MCs with fast
solar wind streams.

% {\S\bf  Roadmap of the paper}\\
In the present paper we {extended the study of} \cite{Gulisano10} to
the outer heliosphere from 1 to 5 AU, analyzing observations made by
the Ulysses spacecraft. We first describe the data used, the method
to define the boundaries, the axis direction of the flux ropes, and
the main characteristics of the velocity profile (Sect.~\ref{Data}).
Next, we study statistically the dependence of the main MC
properties {on} the solar distance (Sect.~\ref{Global}). Then, we
analyze the expansion properties of MCs separately for the perturbed
and non-perturbed cases (Sect.~\ref{Expansion}).  The MCs in
interaction with a strong external magnetic field (such as another
MC) are analyzed separately case by case in
Sect.~\ref{In-interaction}. Finally, we summarize our results and
conclude (Sect.~\ref{Conclusion}).

%%%%%%%%%%%%%%%%%%%%%%%%%%%%%%%%%%%%%%%%%%%%%%%%%%%%%%%%%%%%%%%%%%%%%%%%%%
\begin{figure*}[t!]
\centerline{
\includegraphics[width=0.5\textwidth, clip=]{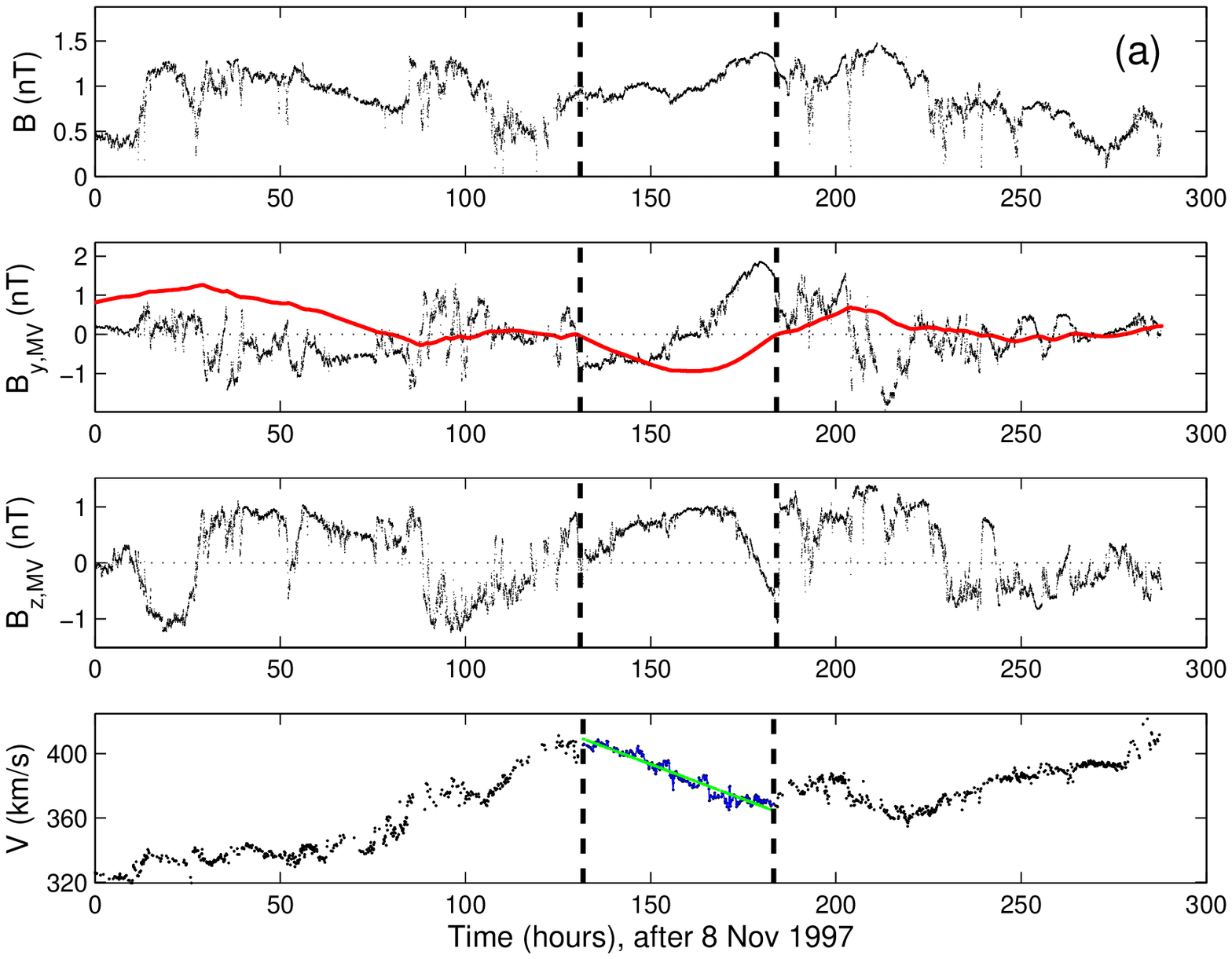}
\includegraphics[width=0.5\textwidth, clip=]{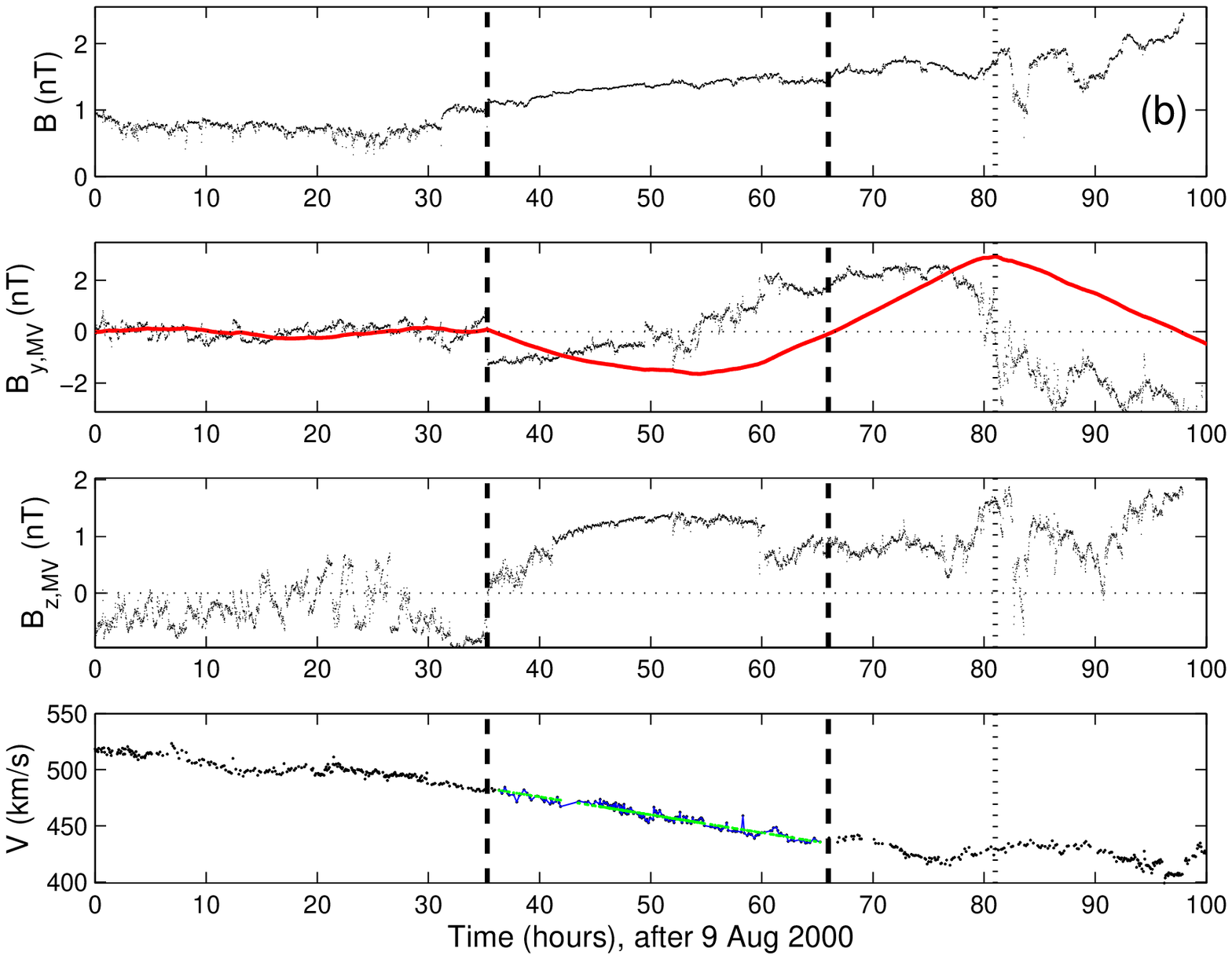}
           }
\caption{Examples of two MCs with velocity profiles {that are} not
significantly perturbed (i.e. that have an almost linear dependence
on time). {\bf (a)} MC 13 travels at low latitude ($2 \degr$)  at
$D=5.3$~AU in a slow SW and {\bf (b)} MC 30 travels at high latitude
($-66 \degr$) at $D=3$~AU in a relatively slow SW for such high
latitude. The vertical dashed lines define the MC (flux rope)
boundaries, and the vertical dotted line defines the rear boundary
of the MC back (Sect.~\ref{D-Boundaries}, a back is only present on
the right panels). The three top panels show the magnetic field norm
and its two main components in the MC frame computed with the MV
method (Sect.~\ref{D-MC-frame}). $B_{\rm z,MV}$ is the MC axial
magnetic field component.  $B_{\rm y,MV}$ is the magnetic field
component both orthogonal to the MC axis and to the radial direction
from the Sun ($\uvec{R}$). The solid red line represents $F_{\rm
y}$, which is the accumulated flux of $B_{\rm y,MV}$
[Eq.~(\ref{Byaccumul})].  The bottom panel shows the observed
velocity component in the radial direction ($\uvec{R}$).  A linear
least squares fit of the velocity (green line) is applied in the
time interval where an almost linear trend is present within the MC.
         }
\label{Fig_not_perturbed}
\end{figure*}
%%%%%%%%%%%%%%%%%%%%%%%%%%%%%%%%%%%%%%%%%%%%%%%%%%%%%%%%%%%%%%%%%%%%%%%%%

%%%%%%%%%%%%%%%%%%%%%%%%%%%%%%%%%%%%%%%%%%%%%%%%%%%%%%%%%%%%%%%%%%%%%%%%%%%%%%%%%%%%%
\section{Data and method} %%%%%%%%%%%%%%%%%%%%%%%%%%%%%
\label{Data}

\subsection{Selected MCs} %%%%%
\label{D-Selected}

% {\S\bf MCs selected}\\
{ We derive the generic expansion properties of MCs in the outer
heliosphere.  We do not select MCs based on their physical
properties (e.g. depending on whether they drive a front shock for
instance), but rather study a set, as complete as possible, of MCs
in a given period of time.}
 We select the MCs previously analyzed by
\citet{Rodriguez04} based on their thermal and energetic properties.
This list of 40 MCs covers the time interval from February 1992 to
August 2002. During the analysis of Ulysses data, we find six extra
MCs during this time interval, which are included at the end of
Table~\ref{Table_MCs}.

% {\S\bf Instruments}\\
Our present study has a different scope from \citet{Rodriguez04} as
our main aim is to quantify the expansion properties of MCs.  We use
data from the Solar Wind
 Observations Over the Poles of the Sun \citep[SWOOPS,][]{Bame92} of
  plasma observations with a temporal cadence of 4 minutes.
The magnetic field data are from the Vector Helium Magnetometer
\citep[VHM,][]{Balogh92},  and have a temporal cadence of one
second.

\subsection{The MC frame } %%%%%
\label{D-MC-frame}
%$\uvec{Z}_{\rm cloud}$
% {\S\bf The RTN frame}\\
  The magnetic and velocity fields observations are in the radial tangential
   normal (RTN) system of reference ($\uvec{R}$, $\uvec{T}$, $\uvec{N}$), where  $\uvec{R}$ points
   from the Sun to the spacecraft, $\uvec{T}$ is the cross product of the Sun's rotation
    vector with $\uvec{R}$, and $\uvec{N}$ completes the right-handed system \citep[e.g., ][]{Fraenz02}.

% {\S\bf The MC frame}\\
  Another very useful system of coordinates is one attached to the local direction of the
  flux rope.  In this MC local frame, $\uvec{Z}_{\rm cloud}$ is {aligned} along the cloud
   axis (with $B_{z, \rm cloud}>0$ in the MC central part), and found by applying the minimum variance (MV) technique to the normalized time series of the
   observed magnetic field \citep[e.g.][and references therein]{Gulisano07}.  We
   define $\uvec{Y}_{\rm cloud}$ in the direction $\uvec{R} \times \uvec{Z}_{\rm cloud}$, and $\uvec{X}_{\rm cloud}$ completes
   the right-handed orthonormal base
($\uvec{X}_{\rm cloud},\uvec{Y}_{\rm cloud},\uvec{Z}_{\rm cloud}$).

% {\S\bf Interest of the MC frame}\\
  In the MC frame, the magnetic field components have a typical behavior that we now describe.  The
  axial field component, $B_{z, \rm cloud}$, typically reaches a maximum in the central part of the MC, and
  declines toward the borders to small, or even negative, values.   The
   component $B_{y, \rm cloud}$ typically changes of sign in the MC central part. When
    large fluctuations are present in some MCs, then $B_{y, \rm cloud}$ has an odd number
    of reversals \citep[e.g.][]{Steed11}.
Finally, the component $B_{x, \rm cloud}$ has its weakest value when
the spacecraft crosses the central
 part of the MC; its relative magnitude provides an indication of the impact parameter \citep{Gulisano07,Demoulin09c}.

% {\S\bf Azimuthal flux conservation}\\
  Another reason for being interested in the MC frame is that it permits us to relate the inward and outward limits of the
  crossed flux rope to the conservation of the azimuthal flux \citep{Dasso06}. We neglect the
  evolution of the magnetic field during the spacecraft crossing period (since the elapsed time is
  small compared to the transit time from the Sun, see \citealp{Demoulin08} for a justification),
  and define the accumulative flux per unit length $L$ along the MC axial
direction as \citep{Dasso07}
    \begin{equation} \label{Byaccumul}
    \frac{F_y(t)}{L}
    = \int_{\tin}^{t} B_{y,cloud}(t') ~V_{x,cloud}(t') ~\rmd t' \,,
    \end{equation}
where $\tin$ is the time of the front boundary.  This time is
selected as a reference since the
 front boundary is usually well-defined, {although} any other reference time can be used without any effect
  on the following conclusions.   Since $B_{y,cloud}$ undergoes its principal change of sign inside a MC, $F_y(t)$, then has
   a main extremum and defines the time called $t_{\rm center}$, which is approximately the time of
   closest approach to the flux rope center.   With the hypothesis of a local symmetry of translation along
   the main axis, the flux surface passing at the position of the spacecraft at time $t_1$ is wrapped around the
    flux rope axis, and is observed at time $t_2$ defined by
    \begin{equation} \label{Fy=0}
    F_y(t_1) = F_y(t_2) \,,
    \end{equation}
from the conservation of the azimuthal flux.  Within the flux rope, this equation relates any
 inward time (time period for inward motion) to its conjugate in the outward (the same but for outward motion).

%%%%%%%%%%%%%%%%%%%%%%%%%%%%%%%%%%%%%%%%%%%%%%%%%%%%%%%%%%%%%%%%%%%%%%%%%%
\begin{figure*}[t!]
\centerline{
\includegraphics[width=0.5\textwidth, clip=]{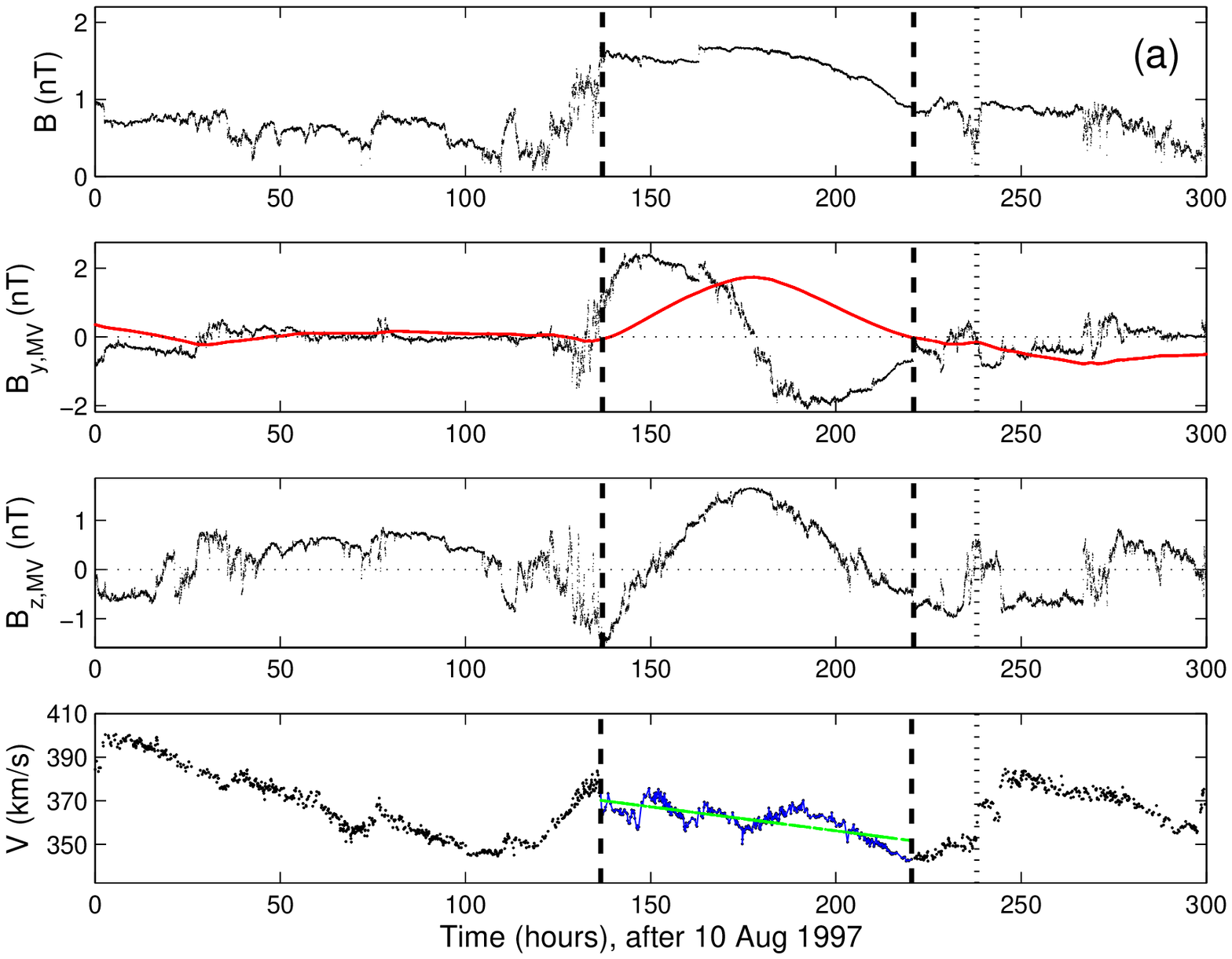}
\includegraphics[width=0.5\textwidth, clip=]{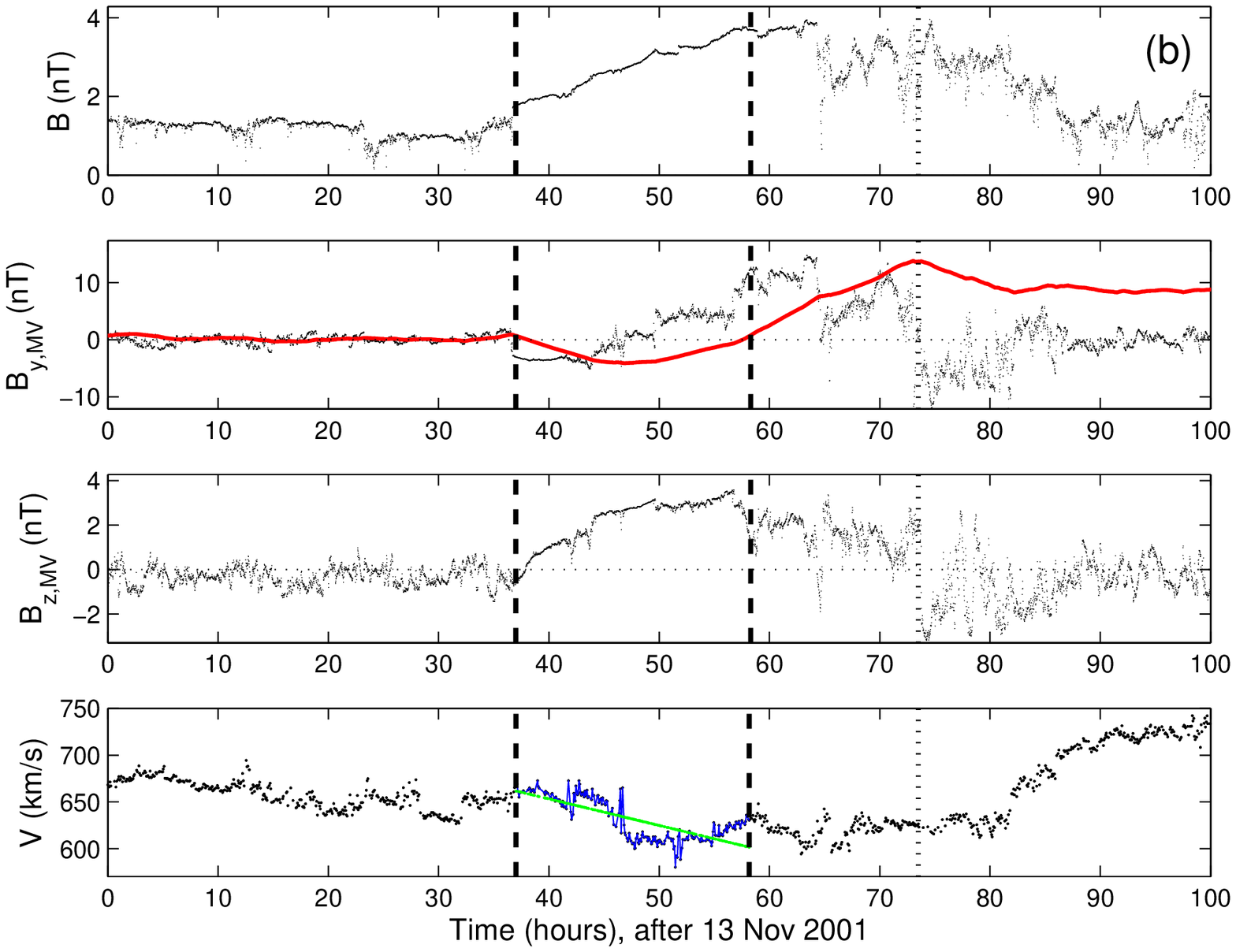}
           }
\caption{Examples of two MCs with perturbed velocity (i.e. deviating
from a linear function of time). {\bf (a)} MC 11 travels at low
latitude ($6 \degr$)  at $D=5.2$~AU in a slow SW and {\bf (b)} MC 36
travels at high latitude ($75 \degr$) at $D=2.3$~AU in a fast SW. We
use the same quantities and drawing convention as
Fig.~\ref{Fig_not_perturbed}. }
 \label{Fig_perturbed}
\end{figure*}
%%%%%%%%%%%%%%%%%%%%%%%%%%%%%%%%%%%%%%%%%%%%%%%%%%%%%%%%%%%%%%%%%%%%%%%%%

\subsection{Definition of the MC boundaries} %%%%%
\label{D-Boundaries}

% {\S\bf How to define the boundaries}\\
   As a first approximation, we choose the MC boundaries from the magnetic field in
    the RTN coordinate system, taking into account the magnetic field properties (see Section~\ref{Introduction}).
We also consider the measured proton temperature and compared them
to the expected temperature in the SW with the same speed
\citep{Lopez86} and to SW temperature accounting for its dependence
on solar distance \citep{Wang05b}. However, a lower temperature than
expected is only used as a guide, since we found that the magnetic
field has sharper changes than the temperature (observed in the time
series of data), and so the magnetic field defines more precisely
the flux rope boundaries.

% {\S\bf Front boundary}\\
   The front, or inward, boundary of a MC is typically the easiest to
    determine.  It is the limit between the fluctuating magnetic field of the sheath
    and the smoother field variation within the flux rope.  A current sheet is
    typically present between these two regions, and the front boundary is
    defined to be there, at a time denoted $\tin$.

% {\S\bf Rear boundary for MCs with back}\\
   The rear, or outward, boundary is more difficult to define,
   particularly because many MCs have a back region with physical properties
   that are in-between those of the MC and SW hence the transition between the flux rope
     and the perturbed SW is frequently not as well defined as at the
     inward boundary. \citet{Dasso06,Dasso07} concluded  that this back region
      is formed, during the transit from the Sun, by reconnection of the flux
      rope with the overtaken magnetic field.  We use the conservation of the
      azimuthal flux between the inward and outward boundary, Eq.~(\ref{Fy=0}), to define
      the outward  boundary.  When a back region is present, the time $\tout$
      of the outward boundary is defined as the first solution of $F_y(\tout)=F_y(\tin)=0$.  However,
      this condition depends on the determination of the MC frame thus an iterative  procedure is needed, as follows.

     % {\S\bf Practical determination of the boundaries}\\
We start with approximate MC boundaries defined in the RTN system,
then perform a MV analysis to find the local frame of the MC. Next,
we analyze the magnetic field components in the local frame, and
redefine the boundaries according to the azimuthal flux conservation
expected in a flux rope [Eq.~(\ref{Fy=0})]. In most cases, the
temporal variations in the magnetic components indicate that the
rear boundary needs to be change its location.  The MV analysis is
then performed only on the flux rope interval.  This is an important
step because an excess of magnetic flux on one side of the flux rope
 typically introduces a systematic bias in the MV directions.
Finally, we perform the same procedure iteratively until convergence
is achieved.  This defines more reliably determined boundaries and
orientation of the flux rope.

 % {\S\bf Result for the rear boundary}\\
 As at the front boundary, we expect to find a current sheet at the outward
  boundary (since it separates two regions with different origins and
  physical properties). A current sheet is typically found at about
  time $\tout$ in most analyzed MCs .  The bending of
  the flux rope axis is one plausible cause of the small difference in
  time between $\tout$ and the closest current sheet.

     % {\S\bf Front boundary for MCs with rear reconnection}\\
In the outward branch of a few MCs, $B_{y, \rm cloud}$ has an
important current sheet and later on it fluctuates before $F_y(t)$
finally vanishes.  This indicates that there is a deficit of
azimuthal flux in the outward branch compared to the inward one. In
this case, $\tout$ is defined by the observed time of the current
sheet, and the flux rope extension is in the time interval $[\tin
,\tout]$, where $\tin $ is redefined by $F_y(\tin )=F_y(\tout)$. The
iterative procedure described above is then applied to improve
$\tin$ and the flux rope orientation.

\subsection{MC groups} %%%%%
\label{D-MC-groups}

% {\S\bf Define perturbed + not-perturbed MCs} \\
Following \citet{Gulisano10}, we divide the data set into three
groups: non-perturbed MCs, perturbed MCs,  and MCs in interaction
(noted {interacting}).

The {interacting} group contains the MCs that have a strong magnetic
field nearby (e.g. another MC). From MHD simulations, their physical
properties are expected to differ significantly from non-interacting
MCs \citep[e.g.,][]{Xiong07}. Observations, indeed confirm that MCs
in the process of interacting could have a different behavior from
non-interacting MCs \citep[e.g.][]{Wang05c,Dasso09}.

The MCs in the {non-perturbed} {and} {perturbed} groups have less
disturbed surroundings than MCs in the {interacting} group. The MCs
in the {non-perturbed} group have an almost linear velocity profile
in more than $75$\% of the flux rope, while in the perturbed MCs
this condition is not satisfied. Two examples of each group are
shown in Figs.~\ref{Fig_not_perturbed},\ref{Fig_perturbed}, the
label (number), the group, and the main properties of each MC are
given in Table~\ref{Table_MCs}.

%%%%%%%%%%%%%%%%%%%%%%%%%%%%%%%%%%%%%%%%%%%%%%%%%%%%%%%%%%%%%%%%%%%%%%%%%%
\begin{figure}[t!]
\includegraphics[width=0.5\textwidth,height=0.3\textwidth, clip=]{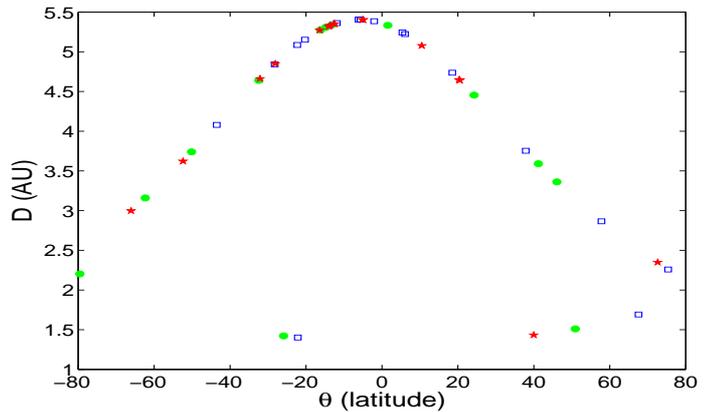}
\caption{Spatial localization of the MCs listed in
Table~\ref{Table_MCs}.  The MCs are separated into three groups:
non-perturbed (filled green circles), perturbed (empty blue
squares), and in the process of interaction (red stars).
}
 \label{Fig_D(lat)}
\end{figure}
%%%%%%%%%%%%%%%%%%%%%%%%%%%%%%%%%%%%%%%%%%%%%%%%%%%%%%%%%%%%%%%%%%%%%%%%%

 %%%%%%%%%%%%%%%%%%% TABLE %%%%%%%%%%%%%%%%%%% TABLE %%%%%%%%%%%%%%%%%%%%%%%%
\begin{table}
\caption{Exponents $e$ in the fit with a power law $D^e$ for $S$,
$<B>$, and $<\Np>$ (found by performing a linear fit in log-log
plots, Fig.~\ref{Fig_lnD}). }
 \label{Table_Exponents}

\begin{center}
\begin{tabular}{@{}cc@{}cc@{}cc@{}cc@{}} % define the column alignment
                                  % l: left, c: center, r: right, @{.} replace the inter-column by a .
    \hline
range of $D$ $^{\mathrm{a}}$ &
\multicolumn{2}{c}{$S$ $^{\mathrm{b}}$}     & \multicolumn{2}{c}{$<B>$ $^{\mathrm{b}}$} &
\multicolumn{2}{c}{$<\Np>$ $^{\mathrm{b}}$} & Ref.$^{\mathrm{c}}$ \\
    \hline

      {\bf all MCs} &&&&&&\\
$[1.4,5.4] $&$ 0.49 $&$ \pm 0.26 $&$-1.20 $&$ \pm 0.20 $&$-1.70 $&$ \pm 0.34 $&\ref{present}\\
    \hline
     {\bf non-perturbed} and \\
      {\bf perturbed MCs} &&&&&&&\\
$[0.3,1]   $&$ 0.78 $&$ \pm 0.12 $&$-1.85 $&$ \pm 0.07 $&$      $&$          $& \ref{Gulisano10}\\
$[1.4,5.4] $&$ 0.56 $&$ \pm 0.34 $&$-1.18 $&$ \pm 0.27 $&$-1.70 $&$ \pm 0.43 $& \ref{present}\\

\hline
       {\bf non-perturbed MCs} &&&&&&&\\
$[0.3,1]   $&$ 0.89 $&$ \pm 0.15 $&$-1.85 $&$ \pm 0.11 $&$      $&$          $& \ref{Gulisano10}\\
$[1.4,5.4] $&$ 0.79 $&$ \pm 0.46 $&$-1.39 $&$ \pm 0.92 $&$-2.24 $&$ \pm 0.66 $& \ref{present}\\

\hline
      {\bf perturbed MCs} &&&&&&&\\
$[0.3,1]   $&$ 0.45 $&$ \pm 0.16 $&$-1.89 $&$ \pm 0.10 $&$      $&$          $& \ref{Gulisano10}\\
$[1.4,5.4] $&$ 0.54 $&$ \pm 0.48 $&$-1.14 $&$ \pm 0.26 $&$-1.40 $&$ \pm 0.50 $& \ref{present}\\

\hline
      {\bf MCs} &&&&&&&\\
$[0.3,1]   $&$      $&$          $&$      $&$          $&$-2.40 $&$ \pm 0.30 $& \ref{Bothmer98}\\
$[0.3,1]   $&$ 1.14 $&$ \pm 0.44 $&$-1.64 $&$ \pm 0.40 $&$-2.44 $&$ \pm 0.46 $& \ref{Leitner07}\\
$[0.3,4.]  $&$ 0.97 $&$ \pm 0.10 $&$-1.80 $&$          $&$-2.8  $&$          $& \ref{Kumar96}\\
$[0.3,4.8] $&$ 0.78 $&$ \pm 0.10 $&$      $&$          $&$      $&$          $& \ref{Bothmer98}\\
$[0.3,5.4] $&$ 0.61 $&$ \pm 0.09 $&$-1.30 $&$ \pm 0.09 $&$-2.62 $&$ \pm 0.07 $& \ref{Leitner07}\\
$[1.4,5.4] $&$      $&$          $&$-0.88 $&$ \pm 0.22 $&$      $&$          $& \ref{Leitner07}\\
\hline
      {\bf ICMEs} &&&&&&&\\
$[0.3,5.4] $&$ 0.61 $&$          $&$-1.52 $&$          $&$      $&$          $&\ref{Wang05b}\\
$[0.3,5.4] $&$ 0.92 $&$ \pm 0.07 $&$-1.40 $&$ \pm 0.08 $&$-2.32 $&$ \pm 0.07 $&\ref{Liu05}\\
\hline

\end{tabular}

   \begin{list}{}{}
   \item[$^{\mathrm{a}}$] The range of solar distances is in astronomical units (AU). In the middle part of the table, MCs are separated in non-perturbed and perturbed  MCs.
   \item[$^{\mathrm{b}}$] The exponents, $e$, are given for the size $S$ in the $\uvec{R}$ direction, the mean magnetic field strength $<B>$, and the mean proton density $<\Np>$.
All quantities are computed within the flux rope boundaries.
   \item[$^{\mathrm{c}}$] We compare our results with our previous results in the inner heliosphere and with other studies of MCs and ICMEs. The results are from:
\refstepcounter{ref} \label{present}    \ref{present}:    present work,
\refstepcounter{ref} \label{Gulisano10} \ref{Gulisano10}: \citet{Gulisano10},
\refstepcounter{ref} \label{Bothmer98}  \ref{Bothmer98}:  \citet{Bothmer98},
\refstepcounter{ref} \label{Leitner07}  \ref{Leitner07}:  \citet{Leitner07},
\refstepcounter{ref} \label{Kumar96}    \ref{Kumar96}:    \citet{Kumar96},
\refstepcounter{ref} \label{Wang05b}    \ref{Wang05b}:    \citet{Wang05b},
\refstepcounter{ref} \label{Liu05}      \ref{Liu05}:      \citet{Liu05}.
   \end{list}

\end{center}
\end{table}
%%%%%%%%%%%%%%%%%%% TABLE %%%%%%%%%%%%%%%%%%% TABLE %%%%%%%%%%%%%%%%%%%%%%%%

%%%%%%%%%%%%%%%%%%%%%%%%%%%%%%%%%%%%%%%%%%%%%%%%%%%%%%%%%%%%%%%%%%%%%%%%%%%%%%%%%%%%%
\section{\p{Magnetic cloud size and mean properties}} %%%%%%%%%%%%%%%%%%%%%%%%%%%%%
%\section{Global MC properties} %%%%%%%%%%%%%%%%%%%%%%%%%%%%%
\label{Global}

\begin{figure}[t!]
\includegraphics[width=0.5\textwidth,height=0.3\textwidth, clip=]{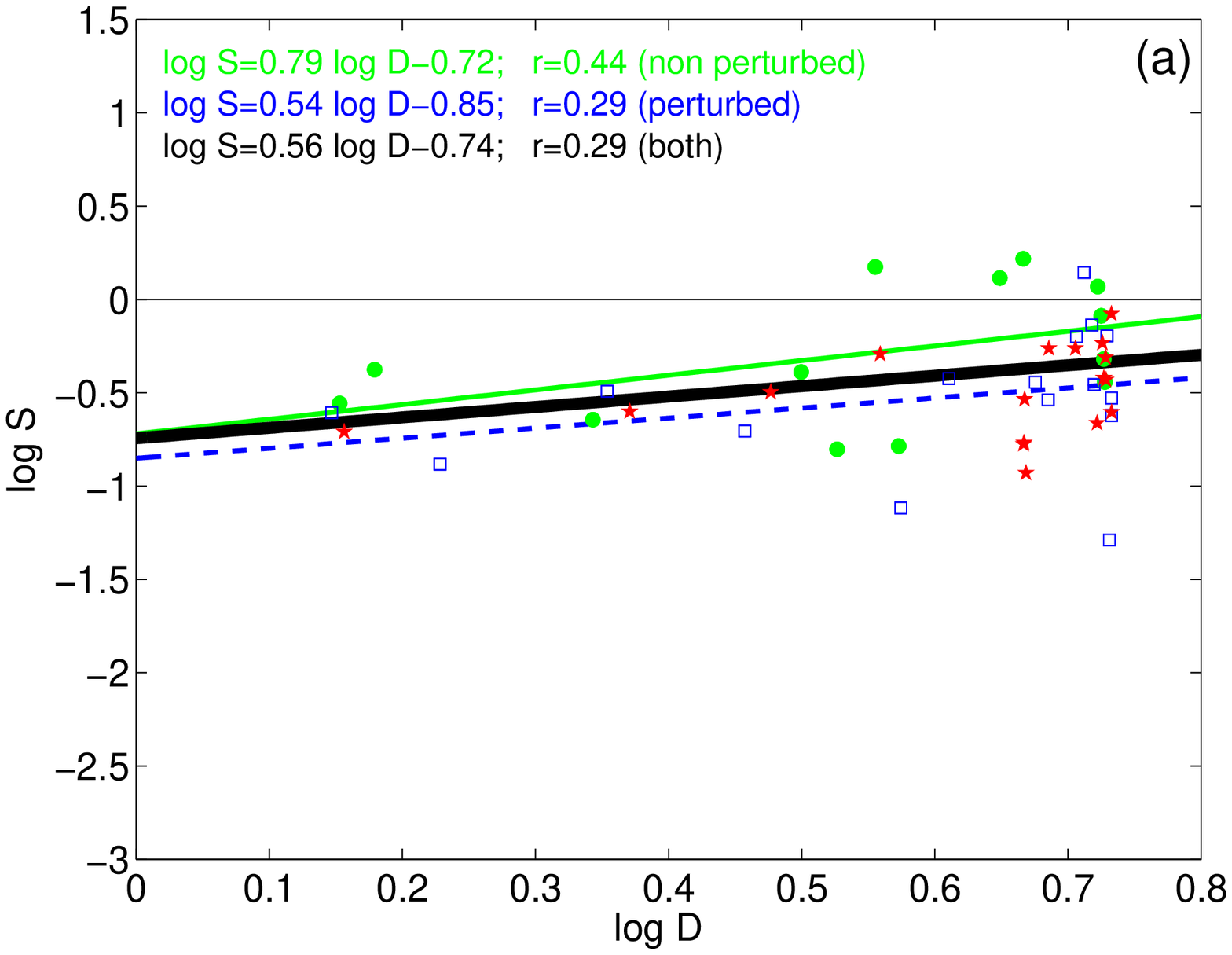}
\includegraphics[width=0.5\textwidth,height=0.3\textwidth, clip=]{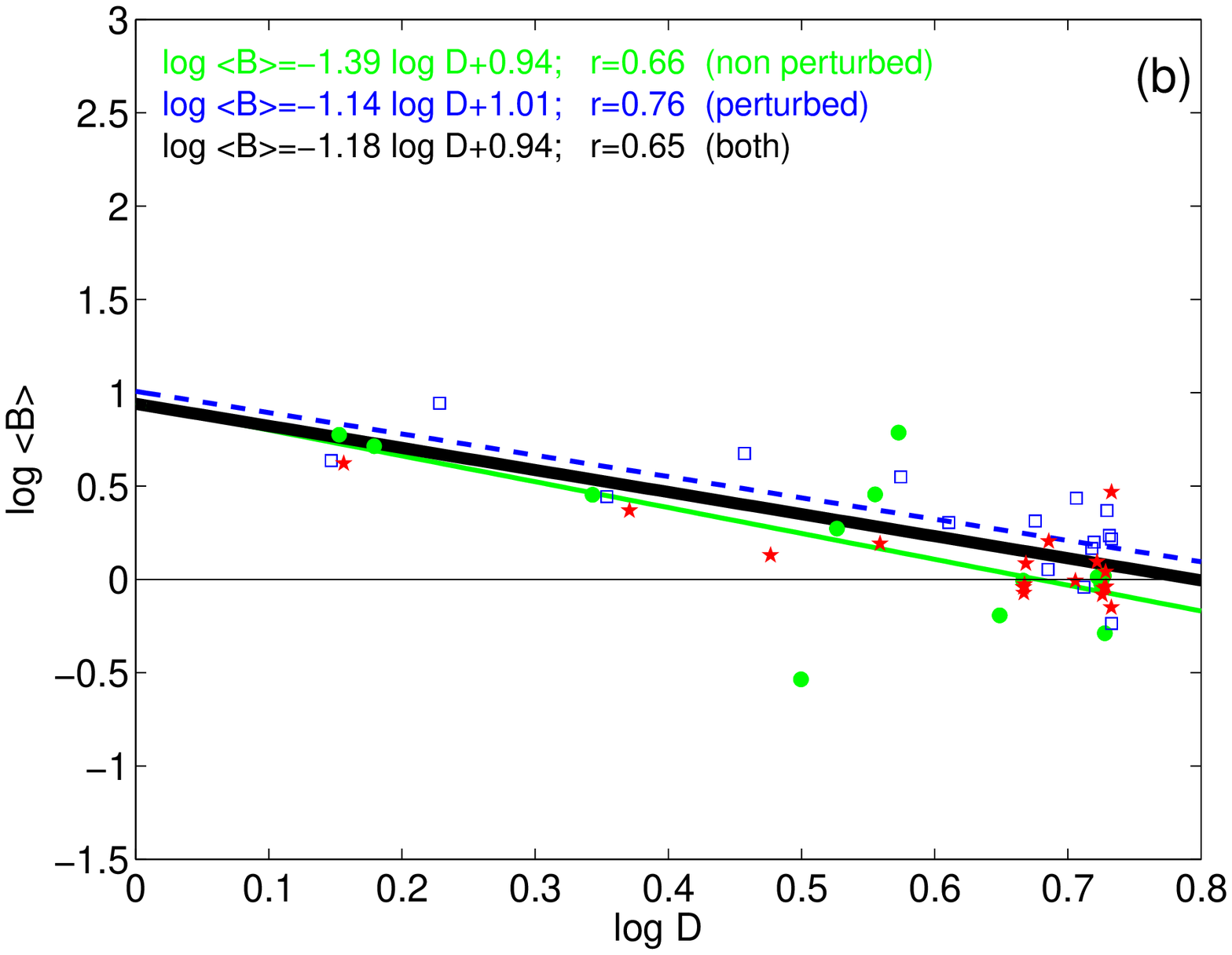}
\includegraphics[width=0.5\textwidth,height=0.3\textwidth, clip=]{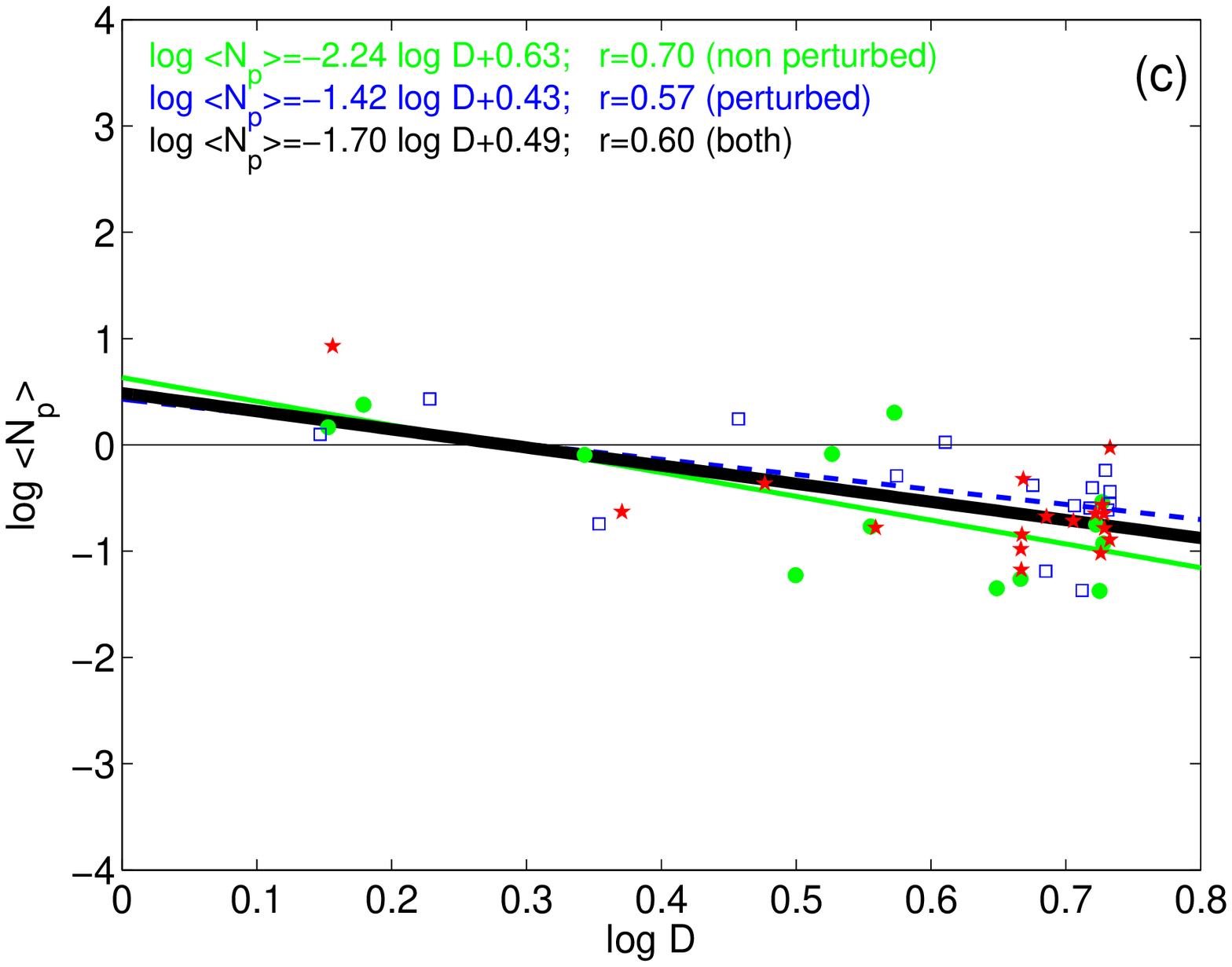}
\caption{ Dependence on the solar distance, $D$ in AU, of
  {\bf (a)} the radial size in the $\uvec{R}$ direction (AU),
  {\bf (b)} the mean magnetic field strength (nT), and
  {\bf (c)} the mean proton density (cm$^{-3}$) with log-log plots.
The MCs are separated in three groups: non-perturbed (filled green
circles), perturbed (empty blue squares), and in interaction (red
stars).  The straight lines are the result of a least squares fit
for non-perturbed (thin continuous green line), perturbed (dashed
blue line), and for  {non-perturbed} {and} {perturbed} MCs (thick
continuous black line). The fits and the absolute value of the
correlation coefficients, r, are added at the top of each panel. }
\label{Fig_lnD}
\end{figure}

\begin{figure}[t!]
\includegraphics[width=0.5\textwidth,height=0.3\textwidth, clip=]{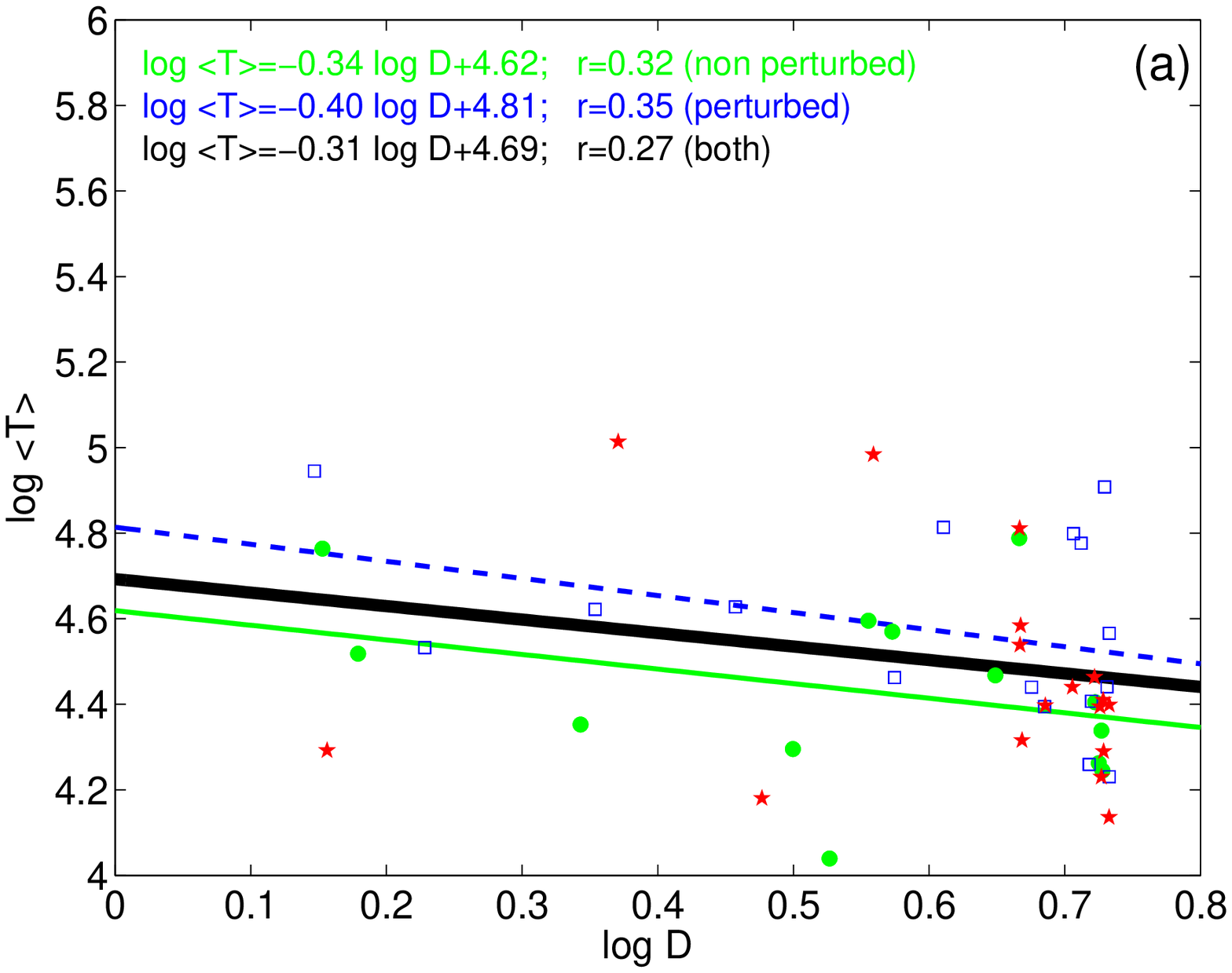}
\includegraphics[width=0.5\textwidth,height=0.3\textwidth, clip=]{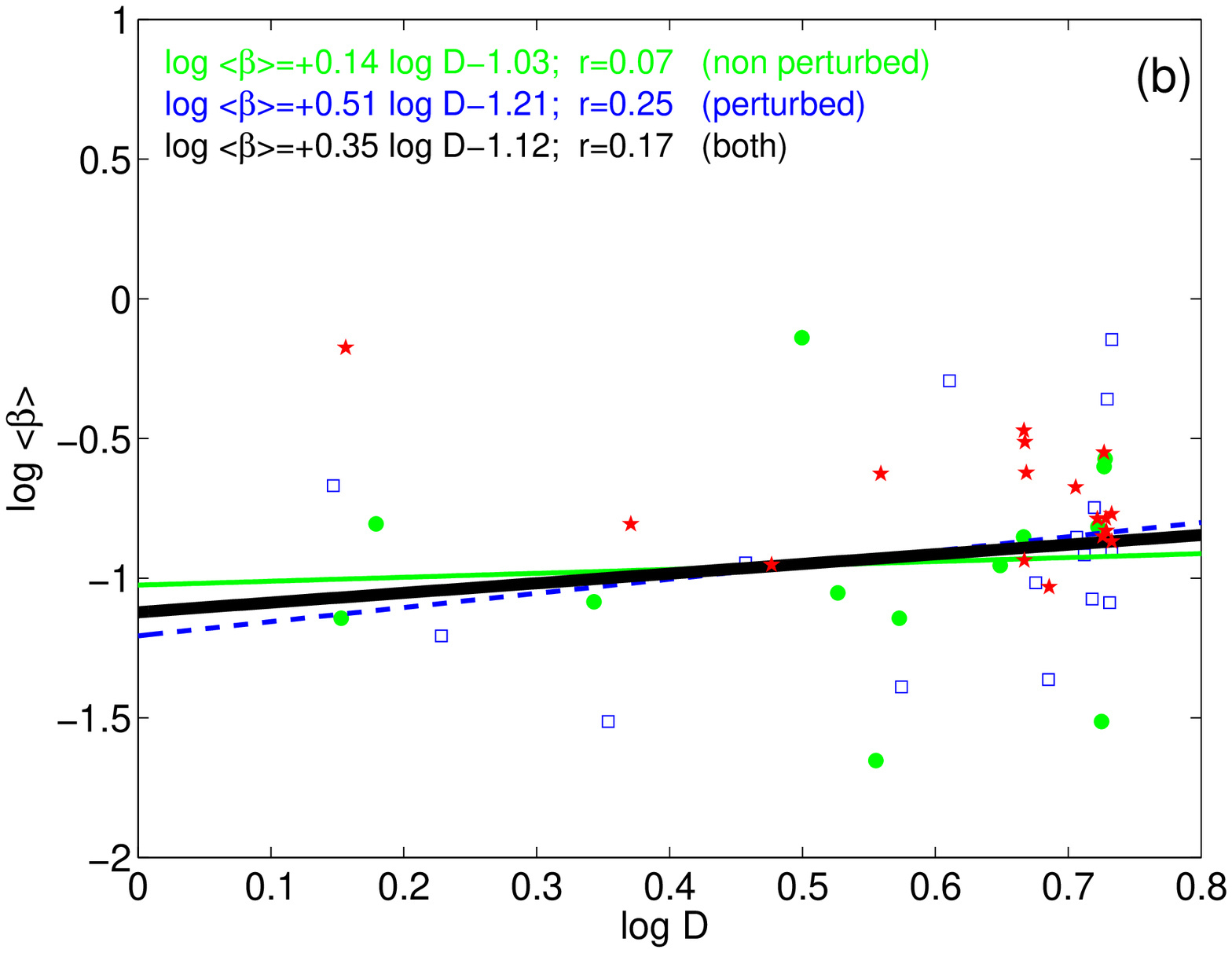}
\caption{Dependence on the solar distance, $D$ in AU, of {\bf (a)}
the mean proton temperature $<T_p>$ (K) and {\bf (b)} the mean
proton $\beta$.  The drawing
 convention is the same than in Fig.~\ref{Fig_lnD}.
}
\label{Fig_plasma}
\end{figure}

  % {\S\bf Aim of the section } \\
We now analyze the \p{size and mean properties, along the spacecraft
trajectory,} of MCs as a function of solar distance $D$. Because of
the trajectory of Ulysses, a correlation between $D$ and the
absolute value of the latitude is present in the analyzed data
(Fig.~\ref{Fig_D(lat)}).
However, the results of the analysis that we now describe do not change significantly
when MCs are grouped in terms of the latitude, so that this correlation
 between $D$ and latitude does not affect our conclusions.  All
\p{the studied} properties are affected by the expansion achieved
from the Sun to Ulysses.  From models \citep[see
e.g.,][]{Chen96,Kumar96,Demoulin09}, all properties are expected to
have a power-law dependence on $D$, we then perform a linear fit for
each property plotted on a log-log scale
(Figs.~\ref{Fig_lnD},\ref{Fig_plasma}), for the {non-perturbed}
 {and} {perturbed} groups. Since MCs in interaction are
strongly case-dependent, their properties are not compared (fitted)
with the global models here.

%%%%%%%%%%%%%%%%%%%%%%%%%%%%%%%%%%%%%%%%%%%%%%%%%%%%%%%%%%%%%%%%%%%%%%%%%%
%%%%%%%%%%%%%%%%%%%%%%%%%%%%%%%%%%%%%%%%%%%%%%%%%%%%%%%%%%%%%%%%%%%%%%%%%

%%%%%%%%%%%%%%%%%%%%%%%%%%%%%%%%%%%%%%%%%%%%%%%%%%%%%%%%%%%%%%%%%%%%%%%%%%
%%%%%%%%%%%%%%%%%%%%%%%%%%%%%%%%%%%%%%%%%%%%%%%%%%%%%%%%%%%%%%%%%%%%%%%%%

\subsection{Magnetic cloud size} %%%%%
\label{G-size}

  % {\S\bf Definition of the size, $S$ } \\
  With the MC boundaries defined in Sect.~\ref{D-Boundaries}, we compute
   the size $S$ in the direction $\uvec{R}$ as the product of the duration
    of the MC and its velocity at closest approach to the center of the
     flux rope.  Compared to most previous works (see Sect.~\ref{Introduction}), this
     typically defines a smaller size as the back region is not included.

  % {\S\bf Dependence $S(D)$. Compare to previous studies } \\
Since the majority of previous studies include all MCs independently
of the groups we defined above, we first compare them to our results
with all MCs included (Table~\ref{Table_Exponents}).  In the range
[1.4,5.4]~AU, we find an exponent that is lower ($0.49\pm 0.26$)
than previous studies on MCs or ICMEs in the inner and outer
heliosphere (lower part of Table~\ref{Table_Exponents}). Taking into
account only {non-perturbed} {and} {perturbed} MCs, the exponent is
slightly higher ($0.56\pm 0.34$), but still lower than previous
studies, including our previous result obtained with Helios data
($0.78\pm 0.12$).

  % {\S\bf Dependence $S(D)$ for P or N MCs} \\
Next, we analyze separately the non-perturbed and perturbed MCs. In
the range [1.4,5.4]~AU, we find comparable exponents to our previous
results obtained with Helios data for both  {non-perturbed} and
perturbed MCs (Table~\ref{Table_Exponents}). {Perturbed MCs} have a
significantly smaller size on average than {non-perturbed} MCs and
this tendency increases with solar distance (Fig.~\ref{Fig_lnD}a).
We find that perturbed MCs  typically have smaller sizes $S$, by a
factor $\approx 1.3$, than non-perturbed MCs.

  % {\S\bf Dependence $S(D)$ for I MCs} \\
The MCs in interaction ({interacting} MCs) have typically smaller
sizes $S$ than the {non-perturbed} MCs (Fig.~\ref{Fig_lnD}a). {We
interpret this result as} a direct effect of the compression and/or
reconnection with the interacting MC or strong magnetic-field
structure.

\subsection{Magnetic field strength} %%%%%
\label{G-<B>}

  % {\S\bf Dependence $<B>(D)$. Compare to previous studies } \\
We define the average field $<B>$ within the flux rope boundaries
defined in Sect.~\ref{D-Boundaries} and proceed as above with $S$.
 The value of $<B>$ decreases far more slowly with distance in the
Ulysses range ($-1.18 \pm 0.27$) than the Helios range ($-1.85 \pm
0.07$) for the {non-perturbed} {and} {perturbed} MCs (a similar
result is obtained when also including the {interacting} MCs).
Finally, combining our Helios and Ulysses results, we have an
intermediate exponent ($-1.5 \pm 0.1$).  This last result is closer
to the results of \citet{Wang05b} and \citet{Liu05} obtained for a
larger set of ICMEs across the same range of distances (bottom of
Table~\ref{Table_Exponents}).

  % {\S\bf Dependence $<B>(D)$ for P or {\bf non perturbed} MCs} \\
The {non-perturbed} MCs have a slightly steeper decrease in magnetic
field strength with distance than the {perturbed} MCs
(Fig.~\ref{Fig_lnD}b).  This result is coherent with the
conservation of magnetic flux combined with a slightly larger
increase in size with distance of the {non-perturbed} MCs compared
to the {perturbed} ones. We also find that the perturbed MCs
typically have a stronger magnetic field, by a factor of between
$\approx 1.4$ and $1.7$, than {non-perturbed} MCs
(Fig.~\ref{Fig_lnD}b), a result that is consistent with the lower
expansion rate for {perturbed} MCs (see Sect.~\ref{G-size}
and~\ref{E-perturbed}).

   % {\S\bf Dependence $<B>(D)$ for I MCs} \\
   The interaction has a weaker effect on $<B>$ than on the size
   $S$:
 %The effect of the interaction on $<B>$ is weaker than for the size $S$.
  There is only a weak tendency for {interacting} MCs to have
  stronger  $<B>$ than {non-perturbed} MCs (Fig.~\ref{Fig_lnD}b).

%%%%%%%%%%%%%%%%%%% TABLE %%%%%%%%%%%%%%%%%%% TABLE %%%%%%%%%%%%%%%%%%%%%%%%
\begin{table*}
\caption{Average and dispersion in the main parameters of MCs from
our present study and \citet{Du10}. }
\label{Table_Average}%%%%%%%%%%%%%%%%%%%%%%%%%%%%%%%%%%%%%%%%%%%%%%%%%%%%%%%%%%%%%%%%%%%%%%%
\begin{center}
\begin{tabular}{c c c c c c c c }
   % define the column alignment:  l: left, c: center, r: right
   % @{.} replace the inter-column by a .  example: c@{ }c| r@{.}l
 \hline
Group $^{\mathrm{a}}$ & $V_c$ $^{\mathrm{b}}$
      & $S_{\rm 1\,AU}$ $^{\mathrm{c}}$
      & $<B>_{\rm 1\,AU}$ $^{\mathrm{c}}$
      & $<\Np>_{\rm 1\,AU}$ $^{\mathrm{c}}$
      & $<\beta_p>$
      & $\gamma$ $^{\mathrm{d}}$
      & $\zeta$ $^{\mathrm{e}}$ \\
      &    km/s    &       AU      &       nT    &  cm$^{-3}$  &               &   \degr   &   \\
 \hline
{\bf all MCs}& 470$\pm$110 & 0.16$\pm$0.12 & 12$\pm$7 & ~9$\pm$8~ & 0.19$\pm$0.17 & 68$\pm$17 & 0.77$\pm$0.75 \\
{\bf non-perturbed} {\bf and} {\bf perturbed}  & 480$\pm$120 & 0.18$\pm$0.14 & 14$\pm$8 & ~9$\pm$8~ & 0.18$\pm$0.19 & 68$\pm$20 & 0.63$\pm$0.59 \\
{\bf non-perturbed}    & 480$\pm$130 & 0.24$\pm$0.16 & 11$\pm$9 &~~7$\pm$10 & 0.17$\pm$0.18 & 63$\pm$24 & 1.05$\pm$0.34 \\
{\bf perturbed}   & 480$\pm$110 & 0.13$\pm$0.09 & 16$\pm$6 & 11$\pm$7~~& 0.19$\pm$0.20 & 72$\pm$15 & 0.28$\pm$0.52 \\
{\bf interacting}    & 470$\pm$130 & 0.12$\pm$0.05 & 10$\pm$6 & ~9$\pm$9~ & 0.22$\pm$0.14 & 66$\pm$13 & 1.00$\pm$0.93 \\
MC$^{\mathrm{f}}$
     & 500$\pm$100 & 0.26$\pm$0.19 & 11$\pm$6 &           &               &           & 0.59$\pm$0.51 \\
non-MC ICME$^{\mathrm{f}}$
     &480$\pm$~90~~& 0.22$\pm$0.21 &~~5$\pm$4 &           &               &           & 0.68$\pm$0.48 \\
 \hline
\end{tabular}
    \begin{list}{}{}
\item[$^{\mathrm{a}}$] The MCs are separated into three groups: non-perturbed, perturbed, and {interacting}.
\item[$^{\mathrm{b}}$ $V_c$ is the velocity at the closest distance from the MC center.]
\item[$^{\mathrm{c}}$] The size and the mean field are normalized to 1~AU according to Eq.~(\ref{S+Bcorrected})
and the proton density with $<\Np>_{\rm 1\,AU} = <\Np> D^{2.2}$,
where $D$ is the solar distance in AU.
\item[$^{\mathrm{d}}$] $\gamma$ is the angle between the MC axis and the radial direction ($\uvec{R}$).
\item[$^{\mathrm{e}}$] $\zeta$ is the nondimensional expansion rate [Eq.~(\ref{zeta})].
\item[$^{\mathrm{f}}$] Computed from the results of \citet{Du10}.
\end{list}

\end{center}
\end{table*}

\subsection{Proton density} %%%%%
\label{G-<Np>}

  % {\S\bf Dependence $<\Np>(D)$. Compare to previous studies } \\
We define the average proton density $<\Np>$ as above for the
magnetic field strength, i.e. within the flux rope boundaries.  For
the {non-perturbed} {and} {perturbed} MCs, the density decreases
significantly less rapidly with distance (exponent $=-1.70 \pm
0.43$) than in previous studies (exponents in the range
$[-2.8,-2.3]$) for both the inner heliosphere and a combination of
both the inner and outer heliosphere. However, we are unaware of
results for the outer heliosphere alone \citep[apart Fig.~10 of][but
no fit is provided]{Leitner07}.

  % {\S\bf Dependence $<\Np>(D)$ for P or N MCs + I MCs} \\
Nevertheless, for the {non-perturbed} MCs the density decreases with
$D$ following a powerlaw that is in closer agreement with previous
studies. In contrast, for perturbed MCs the density decreases much
more slowly with distance. This agrees qualitatively with a lower
expansion of $S$ and a lower decrease in $<B>$ with $D$, while the
difference between {non-perturbed} {and} {perturbed} MCs is more
significant for the average proton density. Even if there is
qualitative agreement, the systematically smaller decrease in
$<\Np>$ with $D$ than in previous studies could be related to the
substantial variability of $\Np$ in MCs \citep[e.g. at 1 AU see
][]{Lepping03} so that $<\Np>$ is case dependent and could be biased
with our relatively small number of MCs. When the same MC is
observed at two distances, this bias is removed. For MC 15, we
computed from the observed data a steeper slope of $2.9$ by using
the mean observed density at both ACE and Ulysses
\citep{Nakwacki11}.

   % {\S\bf Dependence $<\Np>(D)$ for I MCs} \\
 The effect of the interaction on $<\Np>$ is weak, and comparable to the effect on $<B>$:  there is
 only a weak tendency for {interacting} MCs to have
 stronger $<\Np>$ than {non-perturbed} MCs (Fig.~\ref{Fig_lnD}c).  While
  there are too few {interacting} MCs at low $D$ to enable us to draw a strong
  conclusion, the trend of a global decrease in $<\Np>$ with $D$ predominates
   over the effect (typically compression) of an interaction with another MC or a strong B region.

\subsection{Proton temperature} %%%%%
\label{G-T}

  % {\S\bf ~ $<\Tp>(D)$ for N, P and I MCs} \\
   The average proton temperature decreases only weakly with distance (Fig.~\ref{Fig_plasma}a).
The temperatures of the {non-perturbed} {and} {perturbed} MCs have
dependences close to $D^{-0.34}$ and $D^{-0.40}$, respectively. This
contrasts with the decrease found in the SW, which is typically
around $D^{-0.7}$ \citep{Gazis06,Richardson95}. Moreover, the
interacting MCs are not significantly hotter than { non-perturbed}
{and} {perturbed} MCs (apart from two hot {interacting} MCs,
Fig.~\ref{Fig_plasma}a).

  % {\S\bf ~ Compare to \citet{Leitner07}} \\
The above result also differs from the decrease in temperature given
by $\approx D^{-1.6}$ found by \citet{Leitner07} for MCs observed in
the range $[0.3,5.4]$~AU.  However, our results agree with those of
 \citet{Leitner07} since their Fig.~11 shows no decrease in proton
temperature with distance for MCs observed by Ulysses.
We conclude, in agreement with \p{previous results}, %\p{\citet{Leitner07},}
that the proton temperature of MCs in the outer heliosphere
\p{decreases with $D$ much more slowly than in the inner heliosphere
and than} in the SW.

  % {\S\bf ~ Important heating} \\
Moreover, MCs have a much larger volume expansion than the SW, the
main difference being that MCs have a significant expansion in the
radial direction (Table~\ref{Table_Exponents} and
Fig.~\ref{Fig_zeta} below) in contrast to the SW, which has no mean
radial expansion. Considering an adiabatic expansion implies that
MCs should become even cooler than the SW with distance.  The
opposite is in fact observed, which implies that there is a far
greater amount of heating in the MCs than the SW. A significant
source of this heating could be a turbulent cascade to small scales.
Since $\betap << 1$ in MCs (see Sect. \ref{G-beta}), the dissipation
of a small fraction of the magnetic energy provides a large increase
in the plasma temperature, while the same amount of magnetic energy
in the SW provides only a small increase in the plasma temperature
($\betap \approx 1$).

\subsection{Proton plasma $\beta$} %%%%%
\label{G-beta}

  % {\S\bf ~ Prediction of beta decay from decay of T,n,B} \\
The proton plasma $\beta$, denoted $\betap$, is defined as the ratio
of the proton thermal pressure to the magnetic pressure. Thus, if
the decay of the proton density, proton temperature, and magnetic
field intensity are true power laws (respectively, $N_p \propto
D^{-n_n}$, $T_p \propto D^{-n_T}$, and $B \propto D^{-n_B}$), the
decay of $\betap$ will be $\betap \propto D^{-n_n-n_T+2n_B}$, which
in our Ulysses study corresponds to an expected increasing function
with the heliodistance, such as $\beta_{p,exp} \propto
D^{-1.7-0.31+2*1.18} \sim D^{+0.35}$ for the combined sample of {
non-perturbed} {and} {perturbed} MCs
(Figs.~\ref{Fig_lnD}b,c,~\ref{Fig_plasma}a). Another estimation can
be realized by supposing an isotropic expansion of MCs with
distances that increase as $D^m$. Assuming the conservations of both
mass and magnetic flux (i.e., an ideal regime), we theoretically
expect an evolution such that $\betap \propto D^{-3m-n_T+4m}$, which
\p{is $\betap \propto D^{+0.25}$ for {non-perturbed} {and}
{perturbed} MCs and $\betap \propto D^{+0.45}$ for {non-perturbed}
MCs,} and then comparable to the previous estimation.

  % {\S\bf ~ $<\betap>(D)$ for P or N MCs} \\
We compute the mean value of $\betap$ within each studied cloud, finding that  $<\betap>$ \p{is} %\p{results}
lower than unity for all MCs in our sample, and  has a tendency to
slightly increase with solar distance for {non-perturbed} {and}
{perturbed} MCs. We find that $\betap \propto D^{0.35}$ for the
combined groups of MCs (Fig.~\ref{Fig_plasma}b), so a very similar
dependence to that found above assuming exact power laws.

  % {\S\bf ~ Compare to Leitner} \\
  \citet{Leitner07} found a slight decrease, $\approx D^{-0.4}$, for MCs in the range $[0.3,5.4]$~AU.  However, this
   global tendency is not present for MCs observed with Ulysses, as their Fig.~12 shows a tendency of
   increasing $\betap$ with $D$.  We then conclude that $\betap$ has a different dependence on $D$ in both
   the inner and outer heliosphere.

  % {\S\bf ~ Behavior of separated groups} \\
Next, we analyze separately the dependence of $\betap$ on $D$ for
each MC group. The value of $\betap$ is almost independent of $D$
for the {non-perturbed} MCs, and the above increasing trend with
distance can be mainly attributed to the {perturbed} MCs.  The MCs
that are interacting typically have a larger $\betap$ at all $D$
(Fig.~\ref{Fig_plasma}b).

%%%%%%%%%%%%%%%%%%%%%%%%%%%%%%%%%%%%%%%%%%%%%%%%%%%%%%%%%%%%%%%%%%%%%%%%%%
\begin{figure*}[t!]
\centerline{
\includegraphics[width=0.5\textwidth,height=0.3\textwidth, clip=]{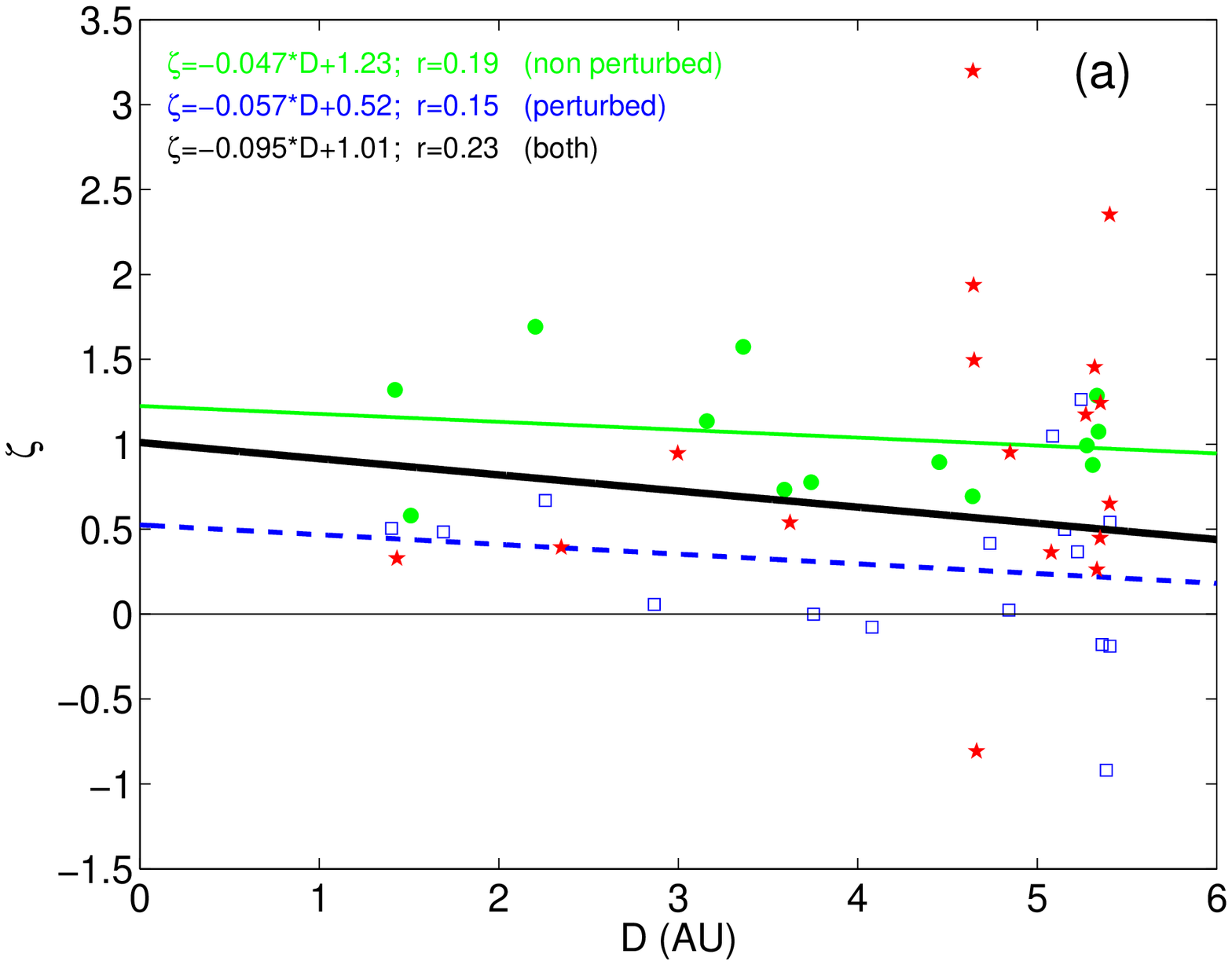}
\includegraphics[width=0.5\textwidth,height=0.3\textwidth, clip=]{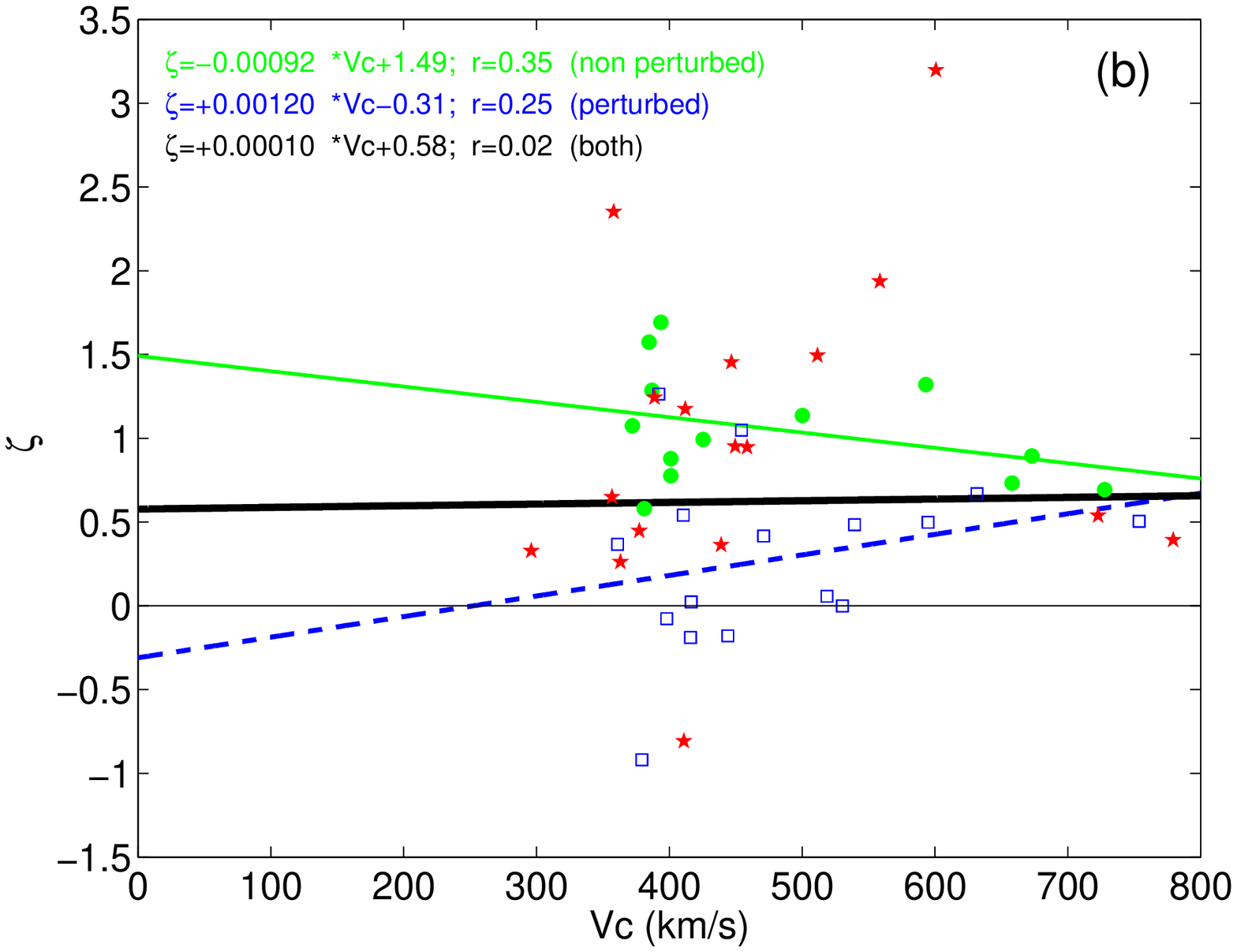}  }
\centerline{
\includegraphics[width=0.5\textwidth,height=0.3\textwidth, clip=]{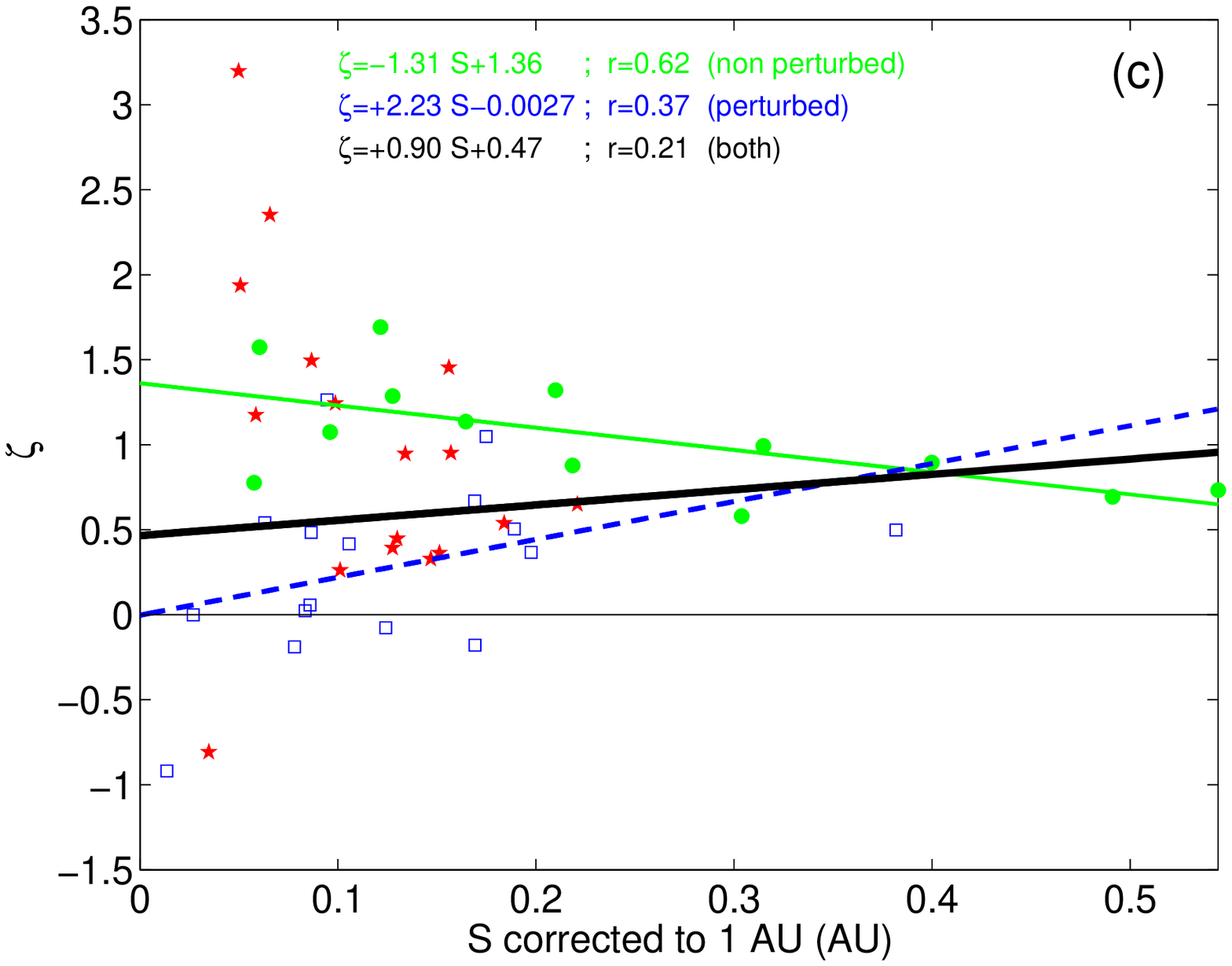}
\includegraphics[width=0.5\textwidth,height=0.3\textwidth, clip=]{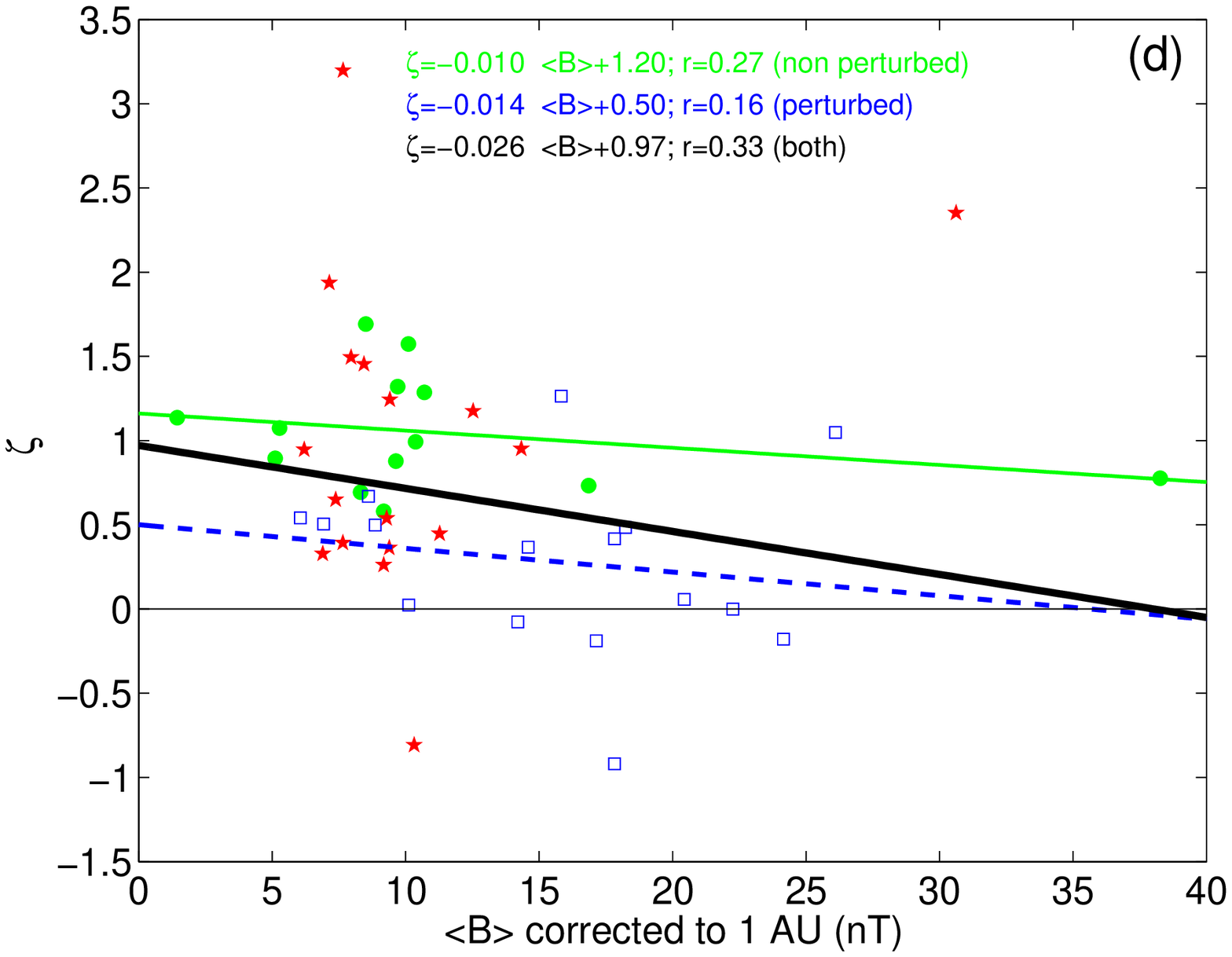}   }
  \caption{The panels {\bf (a-d)} show the correlation analysis that
  tests for the dependence of the non-dimensional expansion
  factor $\zeta$ [Eq.~(\ref{zeta})] on other MC parameters.
The drawing convention is the same as in Fig.~\ref{Fig_lnD}. Both
$S_{\rm 1\,AU}$ and $<B>_{\rm 1\,AU}$ are normalized to 1 AU using
the size and field strength dependence on the distance, according to
 the relationship given in  Figure~\ref{Fig_lnD} for non-perturbed MCs, \p{see Eq.~(\ref{S+Bcorrected}).}
}
 \label{Fig_zeta}
\end{figure*}
%%%%%%%%%%%%%%%%%%%%%%%%%%%%%%%%%%%%%%%%%%%%%%%%%%%%%%%%%%%%%%%%%%%%%%%%%

%%%%%%%%%%%%%%%%%%%%%%%%%%%%%%%%%%%%%%%%%%%%%%%%%%%%%%%%%%%%%%%%%%%%%%%%%%%%%%%%%%%%%
\section{Expansion rate of MCs} %%%%%%%%%%%%%%%%%%%%%%%%%%%%%
\label{Expansion}

% {\S\bf Intro on MC expansion} \\
  Magnetic clouds typically have a velocity profile close to a linear function of time
   with a larger velocity in the front than at the rear. Hence MCs are
    expanding magnetic structures as they move away from the Sun. In
     this section, we characterize their expansion rate.

\subsection{Non-dimensional expansion rate} %%%%%
\label{E-zeta}

% {\S\bf Fit} \\
  The measured temporal profile $V_{R}(t)$ of each MC is fitted using a least
   squares fit with a linear function of time
   \begin{equation}  \label{linear_fit}
   V_{R, \rm fit}(t) = V_{\rm o, fit} + \dVxdt ~t \,,
   \end{equation}
where $\dVxdt$ is the fitted slope.  We always keep the fitting
range within the flux rope, and restricts it to the most linear part
of the observed profile.  The observed linear profile of $V_R$
indicates that different parts of the flux rope expand at about the
same rate in the direction $\uvec{R}$.

% {\S\bf  Define $V_{\rm in}, V_{\rm out}, \Delta V_R$ } \\
The linear fit is used to define the velocities $V_{R, \rm
fit}(t_{in})$ and $V_{R, \rm  fit}(t_{out})$ at the flux rope
boundaries (Sect.~\ref{D-Boundaries}). We then define the full
expansion velocity of a flux rope as
    \begin{equation}  \label{dV}
    \Delta V_R = V_{R, \rm  fit}(t_{in}) - V_{R, \rm  fit}(t_{out}) \,.
    \end{equation}
For non-perturbed MCs, $\Delta V_R$, is very close to the observed
velocity difference  $V_R(t_{in}) - V_R(t_{out})$, see e.g. the
lower panels of Fig.~\ref{Fig_not_perturbed}.  For perturbed MCs,
this procedure minimizes the effects of the perturbations within the
flux rope and especially close to the MC boundaries.

% {\S\bf  Define $\zeta$ } \\
Following  \citet{Demoulin09} and \citet{Gulisano10}, we define the
non-dimensional expansion rate as
   \begin{equation}   \label{zeta}
   \zeta = \frac{\Delta V_R}{\Delta t} \frac{D}{V_c^2} \,
   \end{equation}
which defines a scaling law for the size $S$ of the flux rope (along
$\uvec{R}$) with the distance ($D$) from the Sun as $S \propto
D^{\zeta}$. This simple interpretation of $\zeta$ is obtained with
the following two simplifications which were justified in
\citet{Demoulin09}.  Firstly, we neglect the aging effect (the front
is observed at an earlier time than the rear, so when observed at
the front the flux rope is smaller than when it is observed at the
rear), and secondly, $\zeta$ is approximately constant during the
time interval of the observed flux rope.

% {\S\bf  $\zeta$ estimates the exponent of $S$} \\
While $\zeta$ is a local measurement within the flux rope, it
provides a measure of the exponent $m$ of the flux rope size if it
follows a power law
   \begin{equation}   \label{zeta-Plaw}
   S=S_0 \left(\frac{D}{D_0} \right)^{m}   \,.
   \end{equation}
taking the temporal derivative of $S$, we find that
   \begin{equation}   \label{zeta-m}
   \frac{\rmd S}{\rmd t} = \Delta V_R
                         = \frac{\rmd S}{\rmd D} \frac{\rmd D}{\rmd t}
                   \approx m \frac{S}{D} V_c
                   \approx m \frac{\Delta t V_c^2}{D}  \,.
   \end{equation}
We then find a relation that is equivalent to Eq.~(\ref{zeta}), so
for self-similarly expanding MCs we have $m \approx \zeta$.

\subsection{Expansion of non-perturbed MCs}%%%%%
\label{E-non-perturbed}

% {\S\bf Physics } \\
The main driver of MCs expansion was identified by
\citet{Demoulin09} as the rapid decrease in the total SW pressure
with solar distance . They followed the force-free evolution away
from the Sun of flux ropes with a variety of magnetic field profiles
and assuming either ideal MHD or fully resistive relaxation that
preserves magnetic helicity. Within this theoretical framework, they
demonstrated that a force-free flux rope has an almost self-similar
expansion, so a velocity profile is almost linear with time as
observed by a spacecraft crossing a MC (e.g.
Figs.~\ref{Fig_not_perturbed},\ref{Fig_perturbed}). In the case of a
total SW pressure behaving as $D^{-n_P}$, they also \p{found} that
the normalized expansion rate is $\zeta \approx n_P/4$. These
results apply to a progressive evolution of a flux rope in a quiet
SW, so the non-perturbed MCs are expected to have the closest
properties  to these theoretical results.

% {\S\bf Evolution with distance } \\
   The mean value of $\zeta$ for non-perturbed MCs, observed in
    the range $[1.4,5.4]$ AU, is $<\zeta> \approx 1.05 \pm 0.34$.
This is slightly above the mean values, $\approx 0.80 \pm 0.18$
 and $\approx 0.91 \pm 0.23$, found at 1~AU and in the range
  $[0.3,1.]$ AU, respectively \citep{Demoulin08,Gulisano10}.
This is a small increase compared to the change in $D$,
 which is a factor in the definition of $\zeta$ [Eq.~(\ref{zeta})].
Since $V_c$ has no significant dependence on $D$ \citep{Wang05b},
the increase in $D$ is compensated for mainly by the decrease in
$\Delta V_R/\Delta t$.

% {\S\bf Is the larger mean $<\zeta>$ true ? } \\
The slightly higher mean $<\zeta>$ in the range $[1.4,5.4]$ AU is
mostly due to the higher $\zeta$ values found at smaller $D$ values
(Fig.~\ref{Fig_zeta}a), so in the region that most closely matches
that of both previous studies.  Owing to the characteristics of the
Ulysses orbit, there are indeed only a few observed MCs for
 $D<3$ AU (Fig.~\ref{Fig_D(lat)}) so that the decrease with $D$ in the
 linear fit of $\zeta$ could be due to the specific properties of the
 few detected MCs at these lower distances.   We also recall that
 $D$ is strongly correlated with the heliolatitude $|\theta|$, Fig.~\ref{Fig_D(lat)}, so
 that the dependence $\zeta (D)$ shown in Fig.~\ref{Fig_zeta}a
 could also be an effect of latitude. Nevertheless, we emphasize that the $<\zeta>$ found
here with Ulysses is only slightly higher than the values found
previously at smaller distances from the Sun. These differences are
small compared to the significantly different SW properties of the
slow and fast SWs predominantly present at low and high latitude,
respectively.

% {\S\bf Does $\zeta$ depend on the MC properties?} \\
As in the previous study, we test the possible dependence of $\zeta$ on
 the MC properties.  Both the MC size and field strength depend strongly on
  $D$ (Fig.~\ref{Fig_lnD}a,b).   We use the fits found for the
   non-perturbed MCs to remove, on average, the evolution with $D$.
   Hence we define values at a giving solar distance, here taken at 1~AU,
with the relations
   \begin{eqnarray}
   S_{\rm 1\,AU}   & = &   S ~ D^{-0.79},      \nonumber   \\
   <B>_{\rm 1\,AU} & = & <B> ~ D^{+1.39} \,. \label{S+Bcorrected}
   \end{eqnarray}
The mean MC speed depends only weakly on $D$ in the outer
heliosphere \citep[see e.g. ][for a large set of ICMEs, \p{including
MCs}]{Wang05b}, hence we use below the \p{measured} velocity of the
center $V_{\rm c}$.

% {\S\bf How $\zeta$ depends on the MC properties.} \\
For non-perturbed MCs we find that $\zeta$ is almost independent of
$<B>_{\rm 1\,AU}$ (Fig.~\ref{Fig_zeta}d), as found in the range
$[0.3,1.]$ AU \citep[compared to Fig.~4d of][]{Gulisano10}. A small
difference is that $\zeta$ decreases slightly with $V_{\rm c}$ and
$S_{\rm 1\,AU}$ (Fig.~\ref{Fig_zeta}b,c), while a nearly constant
value was found in the range $[0.3,1.]$ AU \citep[compared to
Fig.~4b,c of][]{Gulisano10}. This difference is probably not due to
a latitude  dependence since the MCs observed at $|\theta |<25\degr$
or $>25\degr$ are evenly distributed in the three plots when we
analyze the two groups independently [not shown], hence similar
results are obtained. Finally, using alternative exponents in
Eq.~(\ref{S+Bcorrected}), in the range given  in
Table~\ref{Table_Exponents}, induces only slight changes in the
linear fits shown in Fig.~\ref{Fig_zeta}, particularly in the
magnitude of the slope.

%%%%%%%%%%%%%%%%%%%%%%%%%%%%%%%%%%%%%%%%%%%%%%%%%%%%%%%%%%%%%%%%%%%%%%%%%%%%%
\begin{figure}[t!]
\centering
\centerline{\includegraphics[width=0.5\textwidth, clip=]{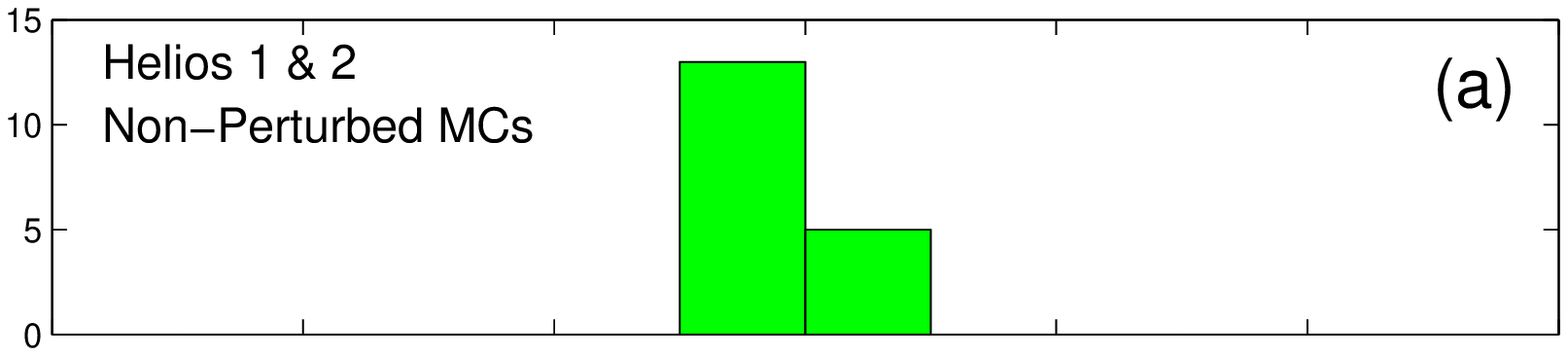}}
\centerline{\includegraphics[width=0.5\textwidth, clip=]{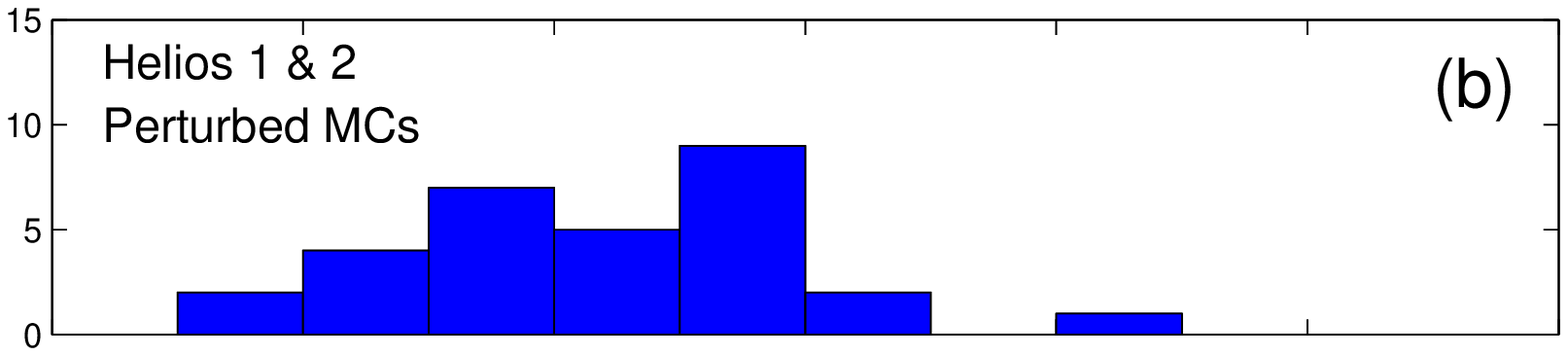}}
\centerline{~\includegraphics[width=0.493\textwidth, clip=]{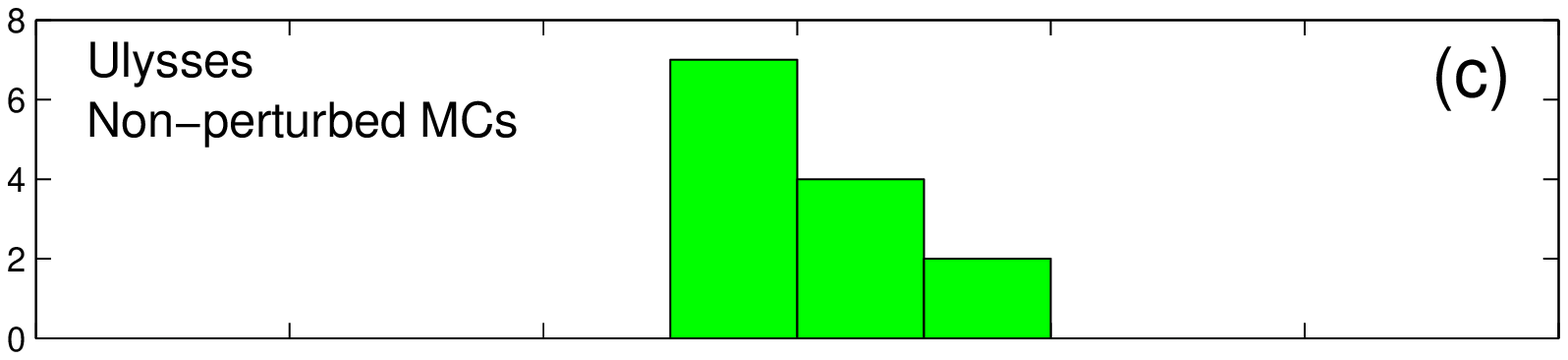}}
\centerline{~\includegraphics[width=0.493\textwidth, clip=]{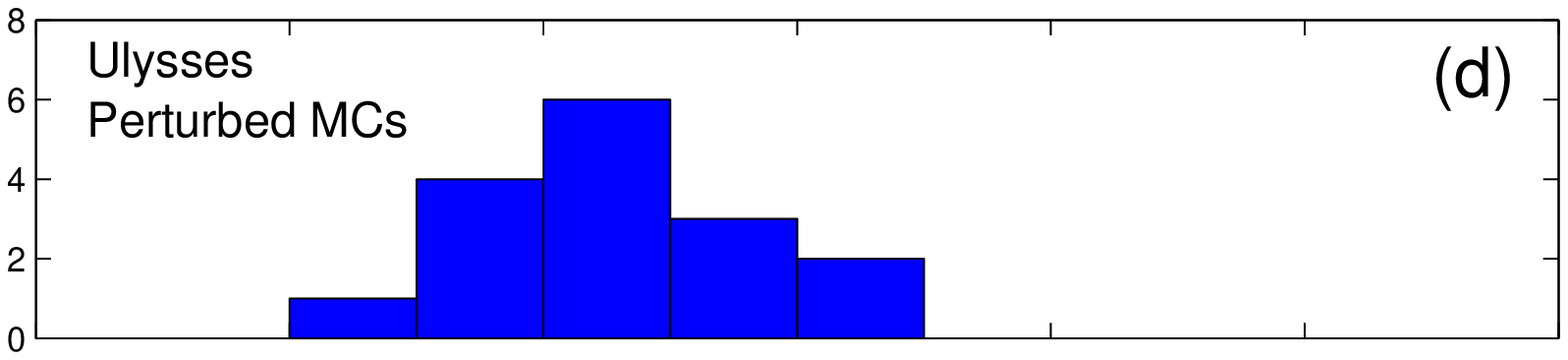}}
\centerline{~\includegraphics[width=0.493\textwidth, clip=]{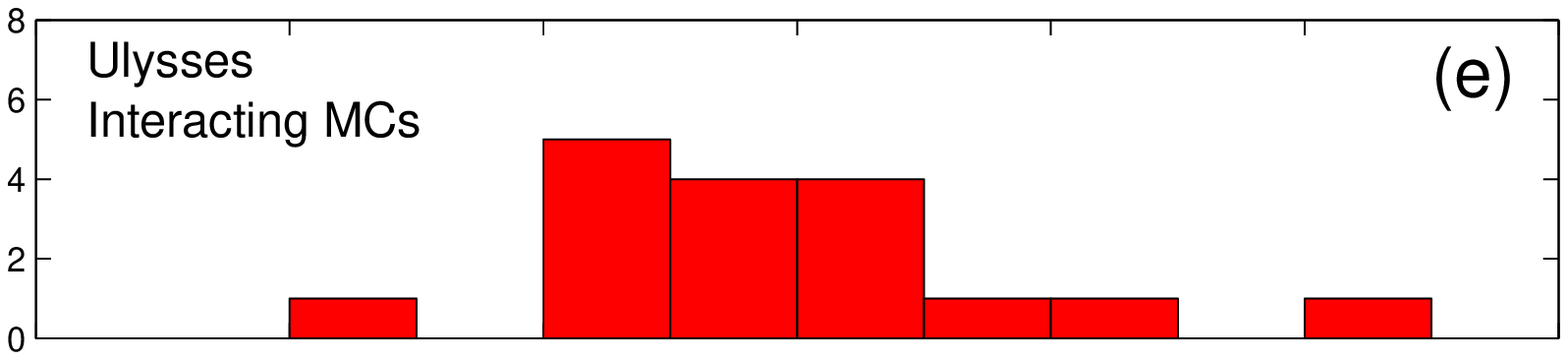}}
\centerline{\includegraphics[width=0.5\textwidth, clip=]{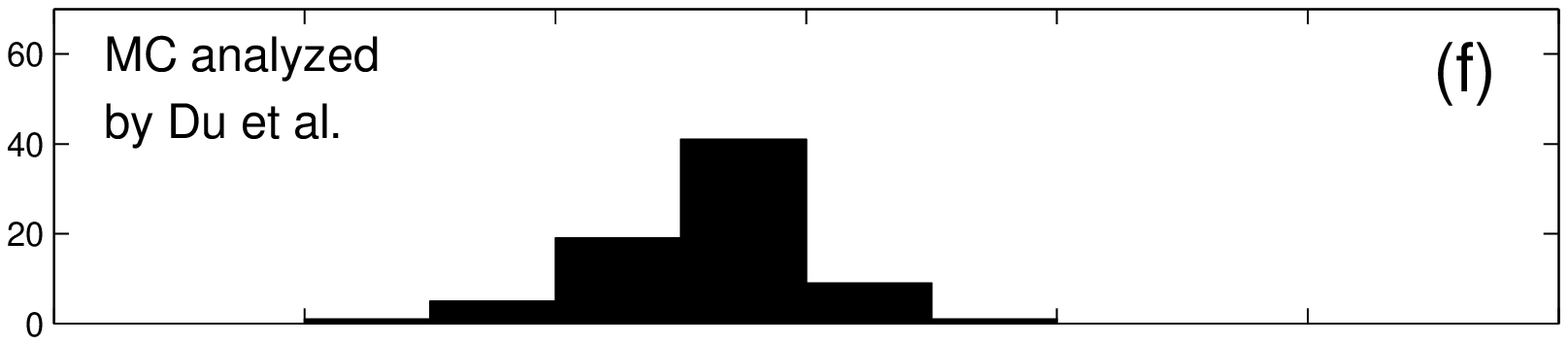}}
\centerline{\includegraphics[width=0.5\textwidth, clip=]{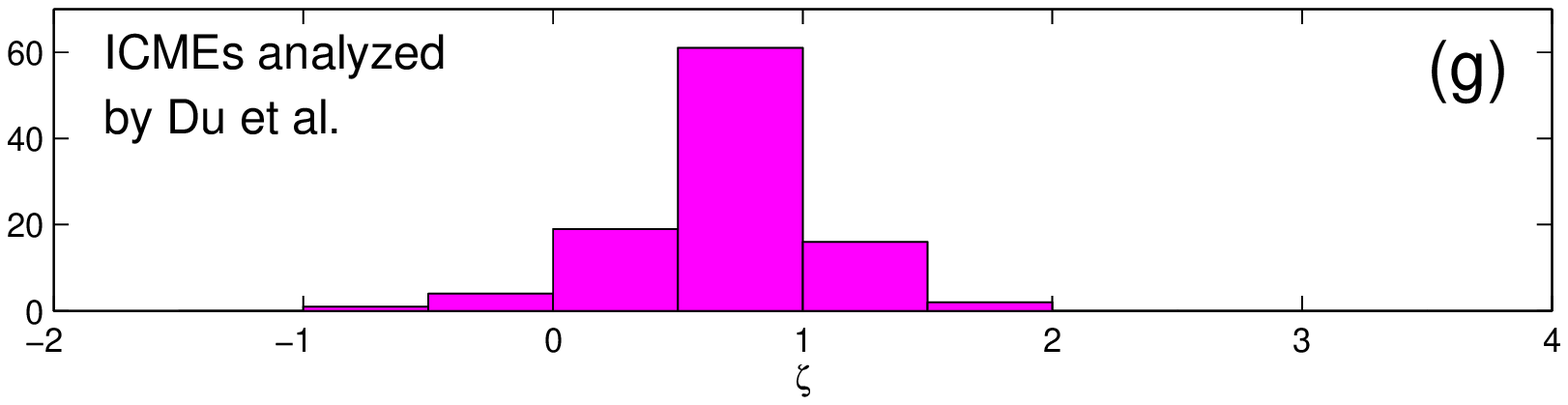}}
\caption{Histograms comparing $\zeta $ of MCs and non-MC ICMEs.
{\bf (a,b)} MCs from Helios spacecraft \citep{Gulisano10}. {\bf (c-e)} MCs from Ulysses
 spacecraft (present work). {\bf (f,g)} MCs and non-MC ICMEs from Ulysses with approximate
  $\zeta $ values computed from the results of \citet{Du10}.
 }
 \label{Fig_histo_zeta}
\end{figure}
%%%%%%%%%%%%%%%%%%%%%%%%%%%%%%%%%%%%%%%%%%%%%%%%%%%%%%%%%%%%%%%%%%%%%%%%%%

%%%%%%%%%%%%%%%%%%%%%%%%%%%%%%%%%%%%%%%%%%%%%%%%%%%%%%%%%%%%%%%%%%%%%%%%%%%%%
\begin{figure}[t!]
\includegraphics[width=0.5\textwidth,height=0.6\textwidth, clip=]{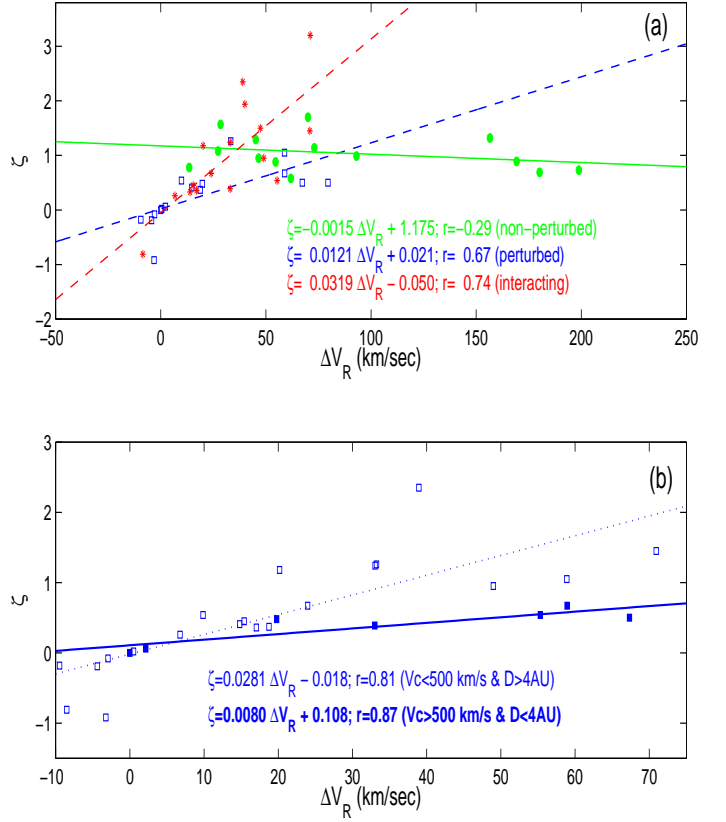}
\caption{Perturbed and non-perturbed MCs have a remarkably different behavior of
 $\zeta $ when they are plotted as a function of $\Delta V_R$. The drawing
  convention is the same as that in Fig.~\ref{Fig_lnD} for panel {\bf (a)}, while
   the perturbed MCs are grouped in terms of velocity and distance in panel {\bf (b)}, as
    indicated by the legend.
}
 \label{Fig_zeta_dV}
\end{figure}
%%%%%%%%%%%%%%%%%%%%%%%%%%%%%%%%%%%%%%%%%%%%%%%%%%%%%%%%%%%%%%%%%%%%%%%%%%

\subsection{Expansion of perturbed MCs} %%%%%
\label{E-perturbed}

% {\S\bf Main difference with non-perturbed MCs} \\
The main difference between perturbed and non-perturbed MCs,
 is that the former has a lower expansion rate
and a larger dispersion, $<\zeta> =0.28\pm 0.52$, compared to
$<\zeta> =1.05\pm 0.34$ for non-perturbed MCs (see also
Fig.~\ref{Fig_histo_zeta}c,d). These results are comparable to the
results found in the distance range $[0.3,1]$~AU by
\citet{Gulisano10} since they found $<\zeta> =0.48\pm 0.79$ and
$<\zeta> =0.91\pm 0.23$ for perturbed and non-perturbed MCs,
respectively (Fig.~\ref{Fig_histo_zeta}a,b). This lower expansion
rate for perturbed MCs is indeed expected from MHD simulations  if
the origin of the perturbation is an overtaking magnetized plasma,
which compresses the MC, hence decreases its expansion rate
\citep{Xiong06,Xiong06b}.

% {\S\bf Over-taking flow: Difference with Helios MCs} \\
A substantial overtaking flow, with a velocity larger than at the
front of the MC, is found in 12 of the \nMCp\ perturbed MCs.  These
strong flows are located behind the observed MC, and typically a
shock, or at least a rapid variation in $V$, is present at the rear
boundary of the back region, as illustrated by MC~11 in
Fig.~\ref{Fig_perturbed}a.  This is the main difference from MCs
observed in the range $[0.3,1]$~AU, as the overtaking flow typically
entered deeply into the perturbed MCs, and moreover, a strong shock
was typically present within the perturbed MC \citep[compare
Fig.~\ref{Fig_perturbed}a with Fig.~2][]{Gulisano10}. This
difference can be interpreted with the MHD simulations of
\citet{Xiong06,Xiong06b}, as follows.  The observations at Helios
distances correspond to the interacting stage when the overtaking
shock travels inside the MC \citep[e.g. see Fig.~5 of][]{Xiong06}.
At larger solar  distances, the overtaking shock exits from the
front of the MC, hence it is not observed inside the MC with Ulysses
data. However, some of the overtaking flow is present behind the MC.
The compression of the overtaking flow thus persists, and most
perturbed MCs are under-expanding.

% {\S\bf Another difference with Helios MCs} \\
Additional evidence of a different evolution stage at Ulysses
compared with Helios is that $\zeta$ shows no significant dependence
on $D$ (Fig.~\ref{Fig_zeta}a), while the mean tendency of $\zeta$
was to increase with $D$ and to reach the value of $<\zeta>$ at 1~AU
of unperturbed MCs for Helios MCs \citep[Fig.~4a][]{Gulisano10}.

% {\S\bf Cases with faster expansion} \\
Only one perturbed MC expands \p{slightly} faster than the mean
expansion rate of unperturbed MCs (Fig.~\ref{Fig_zeta}), while a
stronger over-expansion is present in some MCs in interaction.
These, over-expanding flux ropes can be identified in the above MHD
simulations in the late stage of evolution, as follows. The
overtaking magnetized plasma progressively flows on the MC sides and
overtakes the MC. When only a weak overtaking flow remains at the MC
rear, the expansion rate could increase.  \p{The MC} internal
magnetic pressure is indeed higher (owing to the previously applied
compression) than the pressure value for another MC at the same
position and with no overtaking flow before. \p{This over-pressure
drives a more rapid} MC expansion \citep[e.g. see the summary of two
simulations in Fig.~7 of][]{Xiong06}. Hence, the expansion rate of
perturbed MCs depends on the interaction stage in which they are
observed, \citep[see the cartoon in Figure 6 of][]{Gulisano10}.

% {\S\bf Similarities with Helios MCs: $\zeta(\Delta V)$} \\
Some properties of $\zeta$ \p{for perturbed MCs} are still similar
in \p{the inner and outer heliosphere}. The value of $\zeta$ shows
an increase with both $\Vc$ and $S_{\rm 1\,AU}$
(Fig.~\ref{Fig_zeta}b,c), and a decrease with $<B>_{\rm 1\,AU}$
(Fig.~\ref{Fig_zeta}d) as found with Helios data
\citep[Fig.~4b,c,d][]{Gulisano10}. However, the slope values are
smaller by a factor $\approx 2.2$, and $6$, respectively. Finally,
we note that the above differences are not due to a difference in
the range of parameters since $\Vc$, $S_{\rm 1\,AU}$, and $<B>_{\rm
1\,AU}$ have similar ranges for Helios and Ulysses MCs.

% {\S\bf Analyze of $\zeta(\Delta V)$} \\
   As for Helios MCs, $\zeta$ does not depend on the expansion
    velocity $\Delta V_R$ [defined by Eq.~(\ref{dV})] for non-perturbed
     MCs observed by Ulysses, while $\zeta$ is strongly
correlated with $\Delta V_R$ for perturbed MCs \citep[compare
Fig.~\ref{Fig_zeta_dV}a with Fig.~5 in][]{Gulisano10}.   This result
extends previous Helios results to large distances. Non-perturbed
MCs indeed have an intrinsic expansion rate $\zeta$ (which is given
by the decrease in the total SW pressure with solar distance). This
is not the case for perturbed MCs, which are in a transient stage,
so $\Delta V_R$ cannot be computed by $\rmd S/\rmd t$ as done in
Eq.~(\ref{zeta-m}). Following the derivation of \citet{Gulisano10},
we instead find that
   \begin{equation}   \label{zeta_perturbed}
   \zeta_{\rm perturbed} = \frac{\Delta V_R}{S}\frac{D}{V_c}
               \approx \frac{\Delta V_R D_0^m D^{1-m}}{S_0 V_c} \,,
   \end{equation}
where the MC size $S_0$ is taken at distance $D_0$. For perturbed
MCs, one then expects that $\zeta$ is linearly dependent on $\Delta
V_R$, which is indeed true (Fig.~\ref{Fig_zeta_dV}a). Interacting
MCs are strongly perturbed, and indeed $\zeta$ depends more strongly
on $\Delta V_R$ than the group we call perturbed.

% {\S\bf Detailed analyze of $\zeta(\Delta V)$} \\
  Moreover, from Eq.~(\ref{zeta_perturbed}), a dependence of $\zeta$ on both
   $D$ and $V_c$ is expected.  Within the observed ranges, both
    parameters have a comparable effect on the slope.  Since we
     have a too small number of MCs to perform a test to assess the dependence
     on both $D$ and $V_c$, we only considered the two groups where $D$ and $V_c$
     appear to have a combined effect on the slopes. Hence, these two
      groups have larger slope differences, and luckily, are also the most
       numerous groups.  The first group is defined by $V_c<500$ km/s
        and $D>4$ AU, or more precisely $<V_c>\approx 410$ km/s and $<D>\approx 5.1$ AU.  The
         second group is defined by $V_c>500$ km/s and $D<4$ AU, or more precisely $<V_c>\approx 640$ km/s
          and $<D>\approx 2.6$ AU.  We use a typical $S_0=0.2$ AU value at $D_0=1$ AU, and
           $m=0.54$ as found for perturbed MCs (Table~\ref{Table_Exponents}).
The first group has an expected slope of $0.026$, \p{which is}
comparable to the one \p{derived from the data} ($0.028$,
Fig.~\ref{Fig_zeta_dV}b).  The second group is expected to have a
 smaller slope value of $0.012$ and the observations do indeed give a slope that is even lower by a
factor $1.5$.  One can also compare the ratio of predicted to
observed slopes to eliminate the influence of $S_0$: the observed
ratio of slopes is a factor $1.6$ larger than the predicted one, and
 decreases to $1.4$ if we use $m=0.28$, the mean $\zeta$ value found
for perturbed MCs (Table~\ref{Table_Average}). Taken into account
the uncertainties in the MC parameters, these results indeed show
that Eq.~(\ref{zeta_perturbed}) illustrates that there is a
quantitative dependence of $\zeta$ on both $D$ and $V_c$ \p{for
perturbed MCs.}

\subsection{Global radial expansion} %%%%%
\label{E-Global}

% {\S\bf Expansion of $S$}\\
   The determination of $\zeta$ provides the expansion rate of the
   MC at the time of the observations, while its size $S$ depends on
   the past history of $\zeta$.  If $S$  were observed at another
   solar distance, this would provide global information  on the
    expansion between the two distances (as well as another piece of local
    information about $\zeta$). However, the observation of the same MC
     by two spacecraft is a rare event since it requires a close
      alignment of the spacecraft positions with the propagation
      direction of the MC \citep[one case was analyzed by][]{Nakwacki11}. Nevertheless, the statistical
    evolution of $S$ with $D$ provides a constraint on the mean
    evolution of MCs.  The flux rope size of non-perturbed MCs have a mean dependence close to (Fig.~\ref{Fig_lnD}a)
    \begin{equation} \label{S(D)}
    S \propto D^{0.79 \pm 0.46} \,.
    \end{equation}

% {\S\bf First possibility: reconnection}\\

    If a MC evolves while $\zeta$ remains constant, this implies that $S
\propto D^{\zeta}$ (Sect.~\ref{E-zeta}).  The non-perturbed MCs are
expected to have the most stable $\zeta$ value.  Their
 mean $\zeta$ value, $1.05 \pm 0.34$ (Table~\ref{Table_Average}), indicates that
 they are undergoing a more rapid expansion than indicated by the global results of Eq.~(\ref{S(D)}).   The possible
 difference is that the flux rope size decreases with distance owing to a reconnection with
 the encountered SW magnetic field.   The evidence of this reconnection is a back region observed in a large number of MCs (e.g. a back region is present in the
  three MCs shown in Figs.~\ref{Fig_not_perturbed},\ref{Fig_perturbed}).  If the
  flux rope size were to decrease by a factor $\approx 1.5$ between $1.4$ and $5.4$ following
   a power law of $D$, it would decrease $S$ by $D^{0.3}$, so sufficiently to explain the
   above difference between the mean evolution of $S$ [Eq.~(\ref{S(D)})] and $<\zeta>$ for non-perturbed MCs.

% {\S\bf Second possibility: evolution}\\
An alternative possibility is that, at least, some of the
non-perturbed MCs were in fact previously perturbed so that their
$\zeta$ value was lower.  This would be indicative of a lower global
expansion of the MC, hence a weaker dependence of $S$ on $D$. We
indeed find that $<\zeta> = 0.77 \pm 0.75$ for non-perturbed,
perturbed, and interacting MCs taken together, which is much closer
to the exponent in Eq.~(\ref{S(D)}) than $<\zeta>$ for the
unperturbed MCs alone. According to this view, the unperturbed,
interacting, and perturbed classifications are local ones, which are
only valid around the time of the observation, and a MC changes its
group ({non-perturbed}/{perturbed}/{interacting}) as it travels away
from the Sun. Finally, both of the aforementioned possibilities
could be happening.

\subsection{Magnetic cloud and ICME expansion} %%%%%
\label{E-ICME}

%\s{Luciano suggests to add a paragraph to introduce Table 2 in this section.}

% {\S\bf Summary of Du paper}\\
We now analyze a broader set of MCs and ICMEs from the results
published by \citet{Du10}. They defined ICMEs following
\citet{Wang05b} as the time intervals where the proton temperature,
$T_{p}$, was less than half the expected temperature in the SW with
the same speed when measured in Kelvin degrees. They found that
about 43\% of the identified ICMEs are MCs. Their Ulysses data set
covers the time interval from January 1991 to February 2008.  The
mean values of the principal parameters of this data set are given
in Table~\ref{Table_Average}. Both groups have similar
characteristics to the MCs studied in this paper, with the exception
of a significantly lower magnetic-field strength in non-MC ICMEs.

% {\S\bf Summary of $\Delta V$ results}\\
\citet{Du10} performed a linear fit to the radial velocity through
the full ICME (defined by $T_{p}$, while we restrict ourselves to
the interior of the flux rope in our study). They defined $\Delta V$
as the speed difference between the leading and trailing edges of
the ICME. The distribution of $\Delta V$ is quite similar for MCs
and non-MC ICMEs \citep[Fig.~4]{Du10}.  The main differences are
merely that few MCs have a large $\Delta V$ and that more non-MC
ICMEs have very small $\Delta V$.

% {\S\bf Compare $\zeta$ MC/ICME}\\
In their table of ICMEs, \citet{Du10} provided all the quantities
 to compute $\zeta$ [Eq.~(\ref{zeta})].   We find Gaussian-shaped
  histograms of $\zeta$ for both  MCs and non-MC ICMEs (Fig.~\ref{Fig_histo_zeta}f,g).  They
   are very similar; in particular, their means and dispersions are
    comparable ($\zeta = 0.59 \pm 0.51 $ for MCs and $\zeta = 0.68 \pm 0.48 $ for non-MC ICMEs).
This is compatible with a common idea that non-MC ICMEs have
 the same properties as MCs, but are simply observed \insitu\ near
  their periphery in such a way that the internal flux rope is missed.

% {\S\bf Compare $\zeta$ to our results}\\
 These results are also comparable to our results when we group
  all MCs together, particularly when we consider {non-perturbed
   and perturbed} MCs (Fig.~\ref{Fig_histo_zeta}, Table~\ref{Table_Average}).
However, our detailed analysis permits us to distinguish MCs that
are in different conditions. The MCs in interaction have a broader
range of $\zeta $ (Fig.~\ref{Fig_histo_zeta}e) since they are
strongly affected by the interaction, and they are observed at
different times of the interaction process.   The non-perturbed MCs
logically have the smallest dispersion in $\zeta $
(Fig.~\ref{Fig_histo_zeta}c).  Hence, in addition to the dispersion
in $\zeta $ found from the \citet{Du10} results, we find that
different environments or/and evolution stages are present.

%%%%%%%%%%%%%%%%%%%%%%%%%%%%%%%%%%%%%%%%%%%%%%%%%%%%%%%%%%%%%%%%%%%%%%%%%%%%%%
\section{Magnetic clouds in interaction} %%%%%%%%%%%%%%%%%%%%%%%%%%%%%
\label{In-interaction}

\subsection{Evolution of MCs in a complex solar wind} %%%%%
\label{I-Evolution}

% {\S\bf Structured SW}\\
   Some MCs travel in a structured SW formed by plasma coming from
    different parts of the solar corona.  These plasmas have
    different properties, for instance in terms of plasma speed and magnetic field,
     so they are interacting.  For example, a rarefaction region
      is formed when a fast SW precedes a slow SW, while a
       compression region is formed when a fast SW overtakes
        a slow SW.   Moreover, especially around the maximum
         of the solar cycle, several ICMEs could be ejected
          from the Sun at similar times, \p{then
           interact later on}.   When a MC travels through
           a structured SW, the encountered structures affect its evolution.

% {\S\bf  Generality on MCs in interaction}\\
The group of perturbed MCs, analyzed above, are interpreted as
examples of  MCs affected during their travel from the Sun by the
encountered SW structures. However, since the analyzed data mostly
correspond to events that have interacted in the past, there is no
direct evidence of the interaction in the \insitu\ data. The
previous interactions of these MCs can only be suspected based on an
overtaking flow and a typically lower $\zeta$ value. For some other
MCs, the interaction is ongoing in the observations. In this case,
the MC evolution depends on the type of interaction (e.g. on the
external structure involved, the relative velocity, the magnitude of
its plasma parameters, and the field strength) as well as the time
elapsed since the beginning of the interaction. This implies that
the study of MCs in interaction needs to be done on typically a case
by case basis. We include the MCs that are strongly interacting at
the time of their in situ observation, in a third group ({called
simply interacting}). In the next two subsections, we analyze two
sub-groups (called I$_{\rm B}$ and I$_{\rm MC}$) of these
interacting MCs.

\subsection{MCs with stronger B-field nearby} %%%%%
\label{I-stronger-B}

% {\S\bf  Definition. MCs with a stronger $B$ in the vicinity}\\
In the first sub-group, I$_{\rm B}$, MCs have a strong magnetic
 field in their surroundings, that is typically stronger by a factor
 of above $1.5$ than the  mean field strength inside the MC.
However, this sub-group contains only cases where the nearby
magnetic structure does not have the characteristics of a MC.

% {\S\bf  Analyze the cases. }\\
Magnetic clouds 21 and 46 both have a strong magnetic field within a
large front sheath, which is larger than the MC size. Both have a
low $\zeta$ value ($0.26$ and $0.39$, respectively). The MCs~4 and
27 both have a strong magnetic field behind them within an  extended
region, which is larger than the MC size. Both MCs have  field
strengths that increases from the front to the rear, particularly
MC~27 since its B-profile is unusually increases in an almost linear
way from the front to the rear by a factor of about two. MC~4 has a
relatively small $\zeta$ value ($0.54$), while MC~27 is in
compression ($\zeta=-0.81$). The MCs~10 and 33 also have a strong
magnetic field behind them, while at the difference of both previous
MCs, their internal field strength has several rather unusual
structures, which indicates that they are strongly perturbed. They
both expand less than average as $\zeta=0.36$ and $0.33$,
respectively.

We conclude that the above six MCs are compressed by the surrounding
 strong  magnetic pressure so that they expand less rapidly than the
  average rate (even one MC is in a strong compression stage).   However,
   there is one exception, MC~1 that also has a strong magnetic
    field behind (in a region more extended than the MC size) but
     it is in over-expansion as $\zeta=1.45$.  This MC is likely
      to be in a late interaction stage where the internal pressure,
       build up previously in the under-expansion stage, becomes too
        high and is able to drive an over-expansion (see Sect.~\ref{E-perturbed}).

%%%%%%%%%%%%%%%%%%%%%%%%%%%%%%%%%%%%%%%%%%%%%%%%%%%%%%%%%%%%%%%%%%%%%%%%%%
\begin{figure}[t!]
\includegraphics[width=0.5\textwidth, clip=]{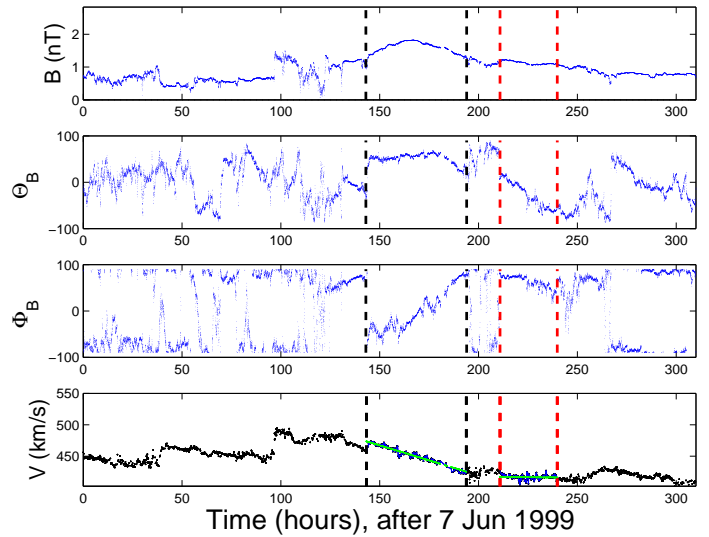}
\caption{ Example of two MCs in interaction. Magnetic cloud 44 is a
large MC with a strong magnetic field; it is interacting with the
smaller MC~26 (the vertical dashed lines define the MC boundaries).
The magnetic field rotation is mostly in the longitude of the field
($\Phi_{B}$) for MC~44 and mostly in the latitude of the field
($\theta_{B}$) for MC~26 wich implies that the MC axis are almost
orthogonal to each other. Nevertheless there is only a weak change
in the magnetic field between the outward branch of MC~44 and the
inward branch of MC~26. The MCs are separated by a sheath-like
region with a magnetic field of different orientation. The bottom
panel shows that MC~26 is not expanding.  A linear least squares fit
of the velocity is indicated by a green line in the time interval
where an almost linear trend is present in each MC.
         }
 \label{Fig_interacting}
\end{figure}
%%%%%%%%%%%%%%%%%%%%%%%%%%%%%%%%%%%%%%%%%%%%%%%%%%%%%%%%%%%%%%%%%%%%%%%%%

\subsection{Interacting MCs} %%%%%
\label{I-Interacting}

% {\S\bf Previous evidence of interacting MCs}\\
  When, after the solar eruption of a magnetic cloud, another
   one is ejected having a more rapid velocity and in a similar direction to
    the first one, an interaction between the two MCs is
     expected near their encounter time.  Depending of the
      relative orientations of the magnetic fields of the
two interacting MCs, magnetic reconnection can develop during
 the interaction.  Previous studies of these processes have
  been performed using observations of individual studied cases
   for both favorable conditions to the magnetic
    reconnection \citep[e.g.,][]{Wang05c}, as well
     as non-reconnection conditions \citep[e.g.,][]{Dasso09}.  The
      dynamical evolution during the interaction of two magnetic
       clouds have been also studied numerically using MHD
        simulations \citep[e.g.,][]{Lugaz05,Xiong07}.  These
         studies have typically found that a linear velocity
          profile is reached in the late evolution phase when
           the two MCs are traveling together.

% {\S\bf Couples of MCs in interaction}\\
  We analyze these interacting MCs, with at least another MC
   in their vicinity, in the second subset of {interacting}
    MCs (I$_{\rm MC}$).  An example is shown in Fig.~\ref{Fig_interacting}.
      The large MC~44 has a typical expanding rate ($\zeta=0.95$), while
       the smaller MC~26 located behind has no  significant expansion
        ($\zeta \approx 0.$).   A second couple of MCs have  similar
         magnetic-field strengths ($<B>\approx 1$~nT), and the front
          MC~18 is in  slight over-expansion ($\zeta=1.24$), while
           behind, MC~19 is in under-expansion ($\zeta=0.45$). A
            third couple of MCs is formed by MCs~15 \p{and} 42. The
            MC~15 was also detected by ACE spacecraft at 1~AU.
              Its expansion rate was classical \citep[$\zeta=0.74$,][]{Nakwacki11}.
               However, when the same MC is measured at 5.4~AU, it is strongly overtaken.
                In particular, its outward branch is compressed ($<B>\approx 1.2$ nT
                 compared to $<B>\approx 0.5$ nT in its inward branch) and in under-expansion
                  ($\zeta \approx 0.57$ compared to $\zeta=0.67$ in its inward branch).
                    This MC~15 is overtaken by MC~42, which has a much stronger field
                     ($<B>\approx 3$~nT) and is in strong over-expansion ($\zeta=2.35$).

% {\S\bf Three MCs in interaction}\\
Finally, the data set contains three MCs traveling close by.  The
three MCs (7,41,8) have similar magnetic field strengths, and this
triad was identified as a single one ICME by \citet{Funsten99} with
an interval of counter-streaming electrons covering the full
duration of the triad. All MCs are in over-expansion ($\zeta =3.2$,
$1.94$, $1.5$, respectively), while the triad taken as a whole has
an expansion rate closer to the mean value of {all the} MCs ($\zeta
\approx 0.8$). Since this last $\zeta$ value is very different than
the $\zeta$ values found within each MC, this triad of MCs is far
from a relaxation state with a linear profile for the full time
period (in contrast to the results of numerical simulations).  At
the same time, the three couples of MCs analyzed above also have
significantly different $\zeta$ values as illustrated for one couple
in Fig.~\ref{Fig_interacting}.

% {\S\bf  Interpretation of simple interaction}\\
 The above three pairs and one triad of interacting MCs represent a
  variety of different cases. Their contrast with the simple evolution
   found in MHD simulations among similar flux ropes, is as follows.
    As with an overtaking fast flow, the MC in front is first
     compressed, so its expansion rate decreases \citep{Lugaz05,Xiong07}.
      The MCs~18 and 19, which have similar field strengths and sizes,
      are the closest to these simulations. However, the following MC is
        in under-expansion while it is the leading MC in \citep{Lugaz05,Xiong07} MHD simulations.

% {\S\bf  Later evolution}\\
When the magnetic reconnection is inhibited between the flux ropes
(by nearly parallel magnetic fields), later on in the MHD
simulations the couple of MCs travel together forming only one
expanding structure, i.e. with a nearly linear velocity profile
across both MCs \citep[see Fig.~4 of][]{Xiong07}.  Such a case was
not found in our data set.

% {\S\bf  More complex interaction.  History.}\\
Other MHD simulations have shown a more complex velocity pattern,
 for example in the case of the interaction of three flux ropes \citep[see Fig.~6 of][]{Lugaz07}.
 We still do not have a complete view of the interaction of MCs of
various sizes, orientations, and magnetic helicities to help us
interpret our observations of interacting MCs. We also have no
information about the history of the interaction, which is crucial
 to improve our understanding of the local measurements since the above MHD simulations
  have shown that the interaction is a time-dependent process (even in the
   simplest interaction cases).
Part of the history can be provided by other data sets such as those
provided by heliospheric imagers and interplanetary radio data for a
more recent time period (with the STEREO spacecraft).

%%%%%%%%%%%%%%%%%%%%%%%%%%%%%%%%%%%%%%%%%%%%%%%%%%%%%%%%%%%%%%%%%%%%%%%%%%%%%%
\section{Summary and conclusions} %%%%%%%%%%%%%%%%%%%%%%%%%%%%%
\label{Conclusion}

% {\S\bf Aims}\\
Our main goal in this paper has been to investigate how the MC
properties, in particular their expansion rates, evolve in the outer
heliosphere based on previous studies of the inner heliosphere and
at 1~AU.  During the travel from the Sun to the location of the
\insitu\ observations, the MC field may become partially reconnected
with the overtaking magnetic field, as indicated by the presence of
a back region in a large fraction of MCs. We have defined the MC
boundaries to retain only the remaining flux rope part in order to
analyze only the region that is not mixed with the SW.

% {\S\bf Grouping}\\
Each observed MC has its own properties, with different temporal
 profiles of its magnetic field and its plasma parameters. Nevertheless we
  have found similarities between MCs that have allowed us to define
   three MC groups.  The MCs within the first group have a radial
    velocity profile that ia almost linear with time, which implies that there is a self-similar
     expansion as theoretically expected, hence this group is
     referred to as non-perturbed (see examples in Fig.~\ref{Fig_not_perturbed}).
In the second group, the temporal velocity profile can deviate
significantly from linearity, hence they are perturbed (see examples
in Fig.~\ref{Fig_perturbed}). Finally, we considered separately the
MCs that have an extensive magnetic-field structure nearby, which
are defined as the group in interaction (one example is shown in
Fig.~\ref{Fig_interacting}).

% {\S\bf Evolution of global parameters}\\
The perturbed MCs have an average size that is smaller than those of
the non-perturbed MCs (Fig.~\ref{Fig_lnD}a), while the opposite
trend is true for the magnetic field strength, proton density, and
temperature (Fig.~\ref{Fig_lnD}b,c,\ref{Fig_plasma}a). All of these
results are consistent with perturbed MCs expanding less with solar
distance than non-perturbed MCs, a property that has indeed been
found independently from analyses of the \insitu\ proton velocity
(Fig.~\ref{Fig_zeta}). For the proton $\beta$, we found no
significant difference between these two MC groups
(Fig.~\ref{Fig_plasma}b).

% {\S\bf Expansion of non-perturbed MCs}\\
The expansion in the radial direction (away from the Sun) is
characterized by the non-dimensional parameter $\zeta$, defined by
Eq.~(\ref{zeta}) from the \insitu\ velocity profile.   When the
observed velocity profile is linear \p{with time},
 the MC size has a power-law expansion such as $S \sim D^\zeta$ [Eq.~(\ref{zeta-Plaw})
  and following text]. For non-perturbed MCs, $\zeta$ has a narrow distribution
   ($\zeta = 1.05\pm 0.34$).  This is slightly larger than the values found in both the
    inner heliosphere ($0.91 \pm 0.23$) and at 1~AU ($0.80 \pm 0.18$).
It is also larger than the global expansion exponent found from the
analysis of the MC size versus helio-distance ($0.79 \pm 0.46$).
This lower exponent for the size could be due to the progressive
pealing of the flux rope by reconnection or/and to a temporal
evolution (Sect.~\ref{E-Global}).

% {\S\bf Expansion of perturbed MCs}\\
The expansion rate of perturbed MCs, $\zeta = 0.28\pm 0.52$, is on
average significantly lower than for non-perturbed MCs, in
quantitative agreement with previous results in the inner
heliosphere. The rate $\zeta$ is also more variable within the group
of perturbed MCs, such as in the inner heliosphere, with even one MC
in over-expansion. These results agree with a temporal evolution of
the expansion rate during the interaction with an overtaking flow as
found in MHD simulations.

% {\S\bf Expansion of MCs in interaction}\\
The MCs in interaction with a stronger magnetic field region have the
 largest variety of expansion rates $\zeta = 1.00\pm 0.93$.
Some of these MCs are in a rapid expansion stage (up to a factor of
three bigger than for non-perturbed MCs), while one MC is even in
compression. This wide variety of expansion rates is partially
present in MHD simulations of interacting flux ropes or a flux rope
overtaken by a fast stream: there is first a compression at the
beginning of the interaction,
 followed by \p{an} increasingly high expansion rate with time that could become an
  over-expansion much latter on (with a magnitude that remains to be quantified).
In all cases the properties of MCs  that area in the process of
interaction are difficult to interpret
 both because the past history is unknown and because they have  a larger variety of properties
 than in MHD simulations.

%%%%%%%%%%%%%%%%%%%%%%%%%%%%%%%%%%%%%%%%%%%%%%%%%%%%%%%%%%%%%%%%%%%%%%%%%%%
%\Online

\begin{appendix}    %First online appendix
\section{List of studied MCs}
\label{Appendix}

% {\S\bf Extra info on MCs}\\
The studied MCs are listed in Table~\ref{Table_MCs}. We added the
 MCs 41 to 46, which were not present in the list of \citet{Rodriguez04}.

%%%%%%%%%%%%%%%%%%% TABLE %%%%%%%%%%%%%%%%%%% TABLE %%%%%%%%%%%%%%%%%%%%%%%%
\begin{table*}
\caption{List of MCs and their main properties.}
\label{Table_MCs}

%%%%%%%%%%%%%%%%%%%%%%%%%%%%%%%%%%%%%%%%%%%%%%%%%%%%%%%%%%%%%%%%%%%%%%%
\begin{center}
\begin{tabular}{rrcrcccccccr@{}r} % define the column alignment
                                  % l: left, c: center, r: right
                                   % @{.} replace the inter-column by a .
                                  % example: c@{ }c| r@{.}l
 \hline
     MC$^{\mathrm{a}}$
   & \multicolumn{1}{c}{$t_c$ $^{\mathrm{b}}$}
   & Group       $^{\mathrm{c}}$
   & $\theta$    $^{\mathrm{d}}$
   & $D$         $^{\mathrm{d}}$
   & $V_c$       $^{\mathrm{e}}$
   & $S$         $^{\mathrm{e}}$
   & $<B>$       $^{\mathrm{f}}$
   & $<\Np>$     $^{\mathrm{f}}$
   & $<\beta_p>$ $^{\mathrm{f}}$
   & $\gamma$    $^{\mathrm{g}}$
   & $\zeta$     $^{\mathrm{h}}$~& \\
   &d-m-y h:m (UT)&     &\degr ~   &AU &km/s &AU &   nT  &cm$^{-3}$&            &  \degr &          & \\
 \hline
%n&       date        time &group&lat.& D   & Vc&   S  &  <B> & <Np> & beta& ga & zeta & \\
 1&17-{\tt Jul}-1992 19:46 &{\bf interacting}&-14 & 5.3 &447& 0.58 & 0.83 & 0.10 & 0.14 & 60 & 1.45 & \\
 2&15-{\tt Nov}-1992 23:18 &{\bf perturbed}&-20 & 5.2 &595& 1.39 & 0.91 & 0.04 & 0.12 & 82 & 0.50 & \\
 3&11-{\tt Jun}-1993 18:29 &{\bf non-perturbed}&-32 & 4.6 &728& 1.65 & 0.98 & 0.05 & 0.14 & 79 & 0.69 &  \\
 4&10-{\tt Feb}-1994 16:46 &{\bf interacting}&-52 & 3.6 &723& 0.51 & 1.55 & 0.17 & 0.24 & 74 & 0.54 &  \\
 5&~3-{\tt Feb}-1995 21:05 &{\bf perturbed}&-22 & 1.4 &754& 0.25 & 4.33 & 1.26 & 0.21 & 55 & 0.50 &  \\
 6&15-{\tt Oct}-1996 01:55 &{\bf non-perturbed}& 24 & 4.6 &673& 1.30 & 0.64 & 0.04 & 0.11 & 67 & 0.89 &  \\
 7&10-{\tt Dec}-1996 20:04 &{\bf interacting}& 21 & 4.6 &601& 0.17 & 0.91 & 0.10 & 0.34 & 48 & 3.20 &  \\
 8&12-{\tt Dec}-1996 22:45 &{\bf interacting}& 20 & 4.7 &512& 0.29 & 0.94 & 0.14 & 0.31 & 65 & 1.50 &  \\
 9&09-{\tt Jan}-1997 09:35 &{\bf perturbed}& 19 & 4.7 &471& 0.36 & 2.05 & 0.42 & 0.10 & 40 & 0.41 &  \\
10&27-{\tt May}-1997 01:52 &{\bf interacting}& 10 & 5.1 &439& 0.55 & 0.98 & 0.19 & 0.21 & 80 & 0.36 &  \\
11&17-{\tt Aug}-1997 09:59 &{\bf perturbed}&  6 & 5.2 &361& 0.73 & 1.47 & 0.25 & 0.08 & 79 & 0.37 &  \\
12&30-{\tt Aug}-1997 19:26 &{\bf perturbed}&  5 & 5.2 &392& 0.35 & 1.58 & 0.40 & 0.18 & 71 & 1.26 &  \\
13&14-{\tt Nov}-1997 19:18 &{\bf non-perturbed}&  2 & 5.3 &387& 0.48 & 1.04 & 0.29 & 0.25 & 80 & 1.29 & \\
14&25-{\tt Jan}-1998 20:24 &{\bf perturbed}& -2 & 5.4 &379& 0.05 & 1.72 & 0.24 & 0.08 & 53 &-0.92 &  \\
15&26-{\tt Mar}-1998 22:53 &{\bf interacting}&-29 & 5.4 &351& 1.10 & 0.48 & 0.12 & 0.37 & 89 & 0.67 &  \\
16&09-{\tt Apr}-1998 20:34 &{\bf perturbed}& -6 & 5.4 &411& 0.24 & 0.58 & 0.36 & 0.71 & 68 & 0.54 &  \\
17&15-{\tt Aug}-1998 02:24 &{\bf perturbed}&-12 & 5.4 &444& 0.64 & 2.34 & 0.58 & 0.44 & 90 &-0.18 &  \\
18&27-{\tt Aug}-1998 00:02 &{\bf interacting}&-13 & 5.4 &389& 0.37 & 0.91 & 0.16 & 0.15 & 73 & 1.24 &  \\
19&30-{\tt Aug}-1998 07:15 &{\bf interacting}&-13 & 5.4 &377& 0.49 & 1.10 & 0.22 & 0.16 & 71 & 0.45 &  \\
20& 8-{\tt Sep}-1998 23:16 &{\bf non-perturbed}&-13 & 5.3 &372& 0.36 & 0.51 & 0.12 & 0.27 & 41 & 1.08 &  \\
21&19-{\tt Sep}-1998 02:19 &{\bf interacting}&-14 & 5.3 &363& 0.38 & 0.90 & 0.27 & 0.28 & 64 & 0.26 &  \\
22& 8-{\tt Oct}-1998 16:35 &{\bf non-perturbed}&-15 & 5.3 &401& 0.82 & 0.95 & 0.04 & 0.03 & 85 & 0.88 &  \\
23& 9-{\tt Nov}-1998 00:53 &{\bf non-perturbed}&-16 & 5.3 &426& 1.17 & 1.03 & 0.18 & 0.15 & 89 & 0.99 &  \\
24&12-{\tt Nov}-1998 16:16 &{\bf interacting}&-16 & 5.3 &412& 0.22 & 1.24 & 0.23 & 0.16 & 60 & 1.18 &  \\
25& 5-{\tt Mar}-1999 01:18 &{\bf perturbed}&-22 & 5.1 &454& 0.63 & 2.72 & 0.27 & 0.14 & 86 & 1.05 &  \\
26&16-{\tt Jun}-1999 08:22 &{\bf perturbed}&-28 & 4.8 &417& 0.29 & 1.13 & 0.65 & 0.04 & 88 & 0.02 &  \\
27&17-{\tt Aug}-1999 06:46 &{\bf interacting}&-32 & 4.7 &411& 0.12 & 1.21 & 0.48 & 0.24 & 74 &-0.81 &  \\
28&31-{\tt Mar}-2000 21:32 &{\bf non-perturbed}&-50 & 3.7 &401& 0.16 & 6.12 & 2.01 & 0.07 & 73 & 0.78 &  \\
29&15-{\tt Jul}-2000 14:52 &{\bf non-perturbed}&-62 & 3.2 &500& 0.41 & 0.29 & 0.06 & 0.73 & 21 & 1.14 &  \\
30&11-{\tt Aug}-2000 06:22 &{\bf non-perturbed}&-66 & 3.0 &459& 0.32 & 1.35 & 0.43 & 0.11 & 85 & 0.95 &  \\
31& 6-{\tt Dec}-2000 23:15 &{\bf non-perturbed}&-80 & 2.2 &394& 0.23 & 2.84 & 0.80 & 0.08 & 31 & 1.70 &  \\
32&12-{\tt Apr}-2001 01:28 &{\bf non-perturbed}&-26 & 1.4 &593& 0.28 & 5.95 & 1.47 & 0.07 & 87 & 1.32 &  \\
33& 7-{\tt Jul}-2001 13:35 &{\bf interacting}& 40 & 1.4 &296& 0.20 & 4.18 & 8.51 & 0.67 & 81 & 0.33 &  \\
34&24-{\tt Jul}-2001 04:36 &{\bf non-perturbed}& 51 & 1.5 &381& 0.42 & 5.17 & 2.39 & 0.16 & 30 & 0.58 &  \\
35&24-{\tt Aug}-2001 19:09 &{\bf perturbed}& 68 & 1.7 &539& 0.13 & 8.78 & 2.70 & 0.06 & 88 & 0.48 &  \\
36&14-{\tt Nov}-2001 22:53 &{\bf perturbed}& 75 & 2.3 &632& 0.32 & 2.77 & 0.18 & 0.03 & 79 & 0.67 &  \\
37&12-{\tt Feb}-2002 13:32 &{\bf perturbed}& 58 & 2.9 &519& 0.20 & 4.73 & 1.76 & 0.11 & 58 & 0.06 &  \\
38& 5-{\tt May}-2002 15:02 &{\bf non-perturbed}& 46 & 3.4 &385& 0.16 & 1.87 & 0.82 & 0.09 & 70 & 1.57 &  \\
39&16-{\tt Jun}-2002 21:19 &{\bf non-perturbed}& 41 & 3.6 &658& 1.49 & 2.85 & 0.17 & 0.02 & 70 & 0.73 &  \\
40&18-{\tt Jul}-2002 06:51 &{\bf perturbed}& 38 & 3.8 &530& 0.08 & 3.54 & 0.51 & 0.04 & 79 & 0.00 &  \\ %-0.0003
41&11-{\tt Dec}-1996 15:18 &{\bf interacting}& 20 & 4.6 &558& 0.17 & 0.85 & 0.07 & 0.12 & 55 & 1.94 &  \\
42&29-{\tt Mar}-1998 07:06 &{\bf interacting}& -5 & 5.4 &358& 0.25 & 2.93 & 0.94 & 0.14 & 59 & 2.35 &  \\
43&20-{\tt Apr}-1998 12:38 &{\bf perturbed}& -6 & 5.4 &416& 0.30 & 1.64 & 0.28 & 0.13 & 65 &-0.19 &  \\
44&14-{\tt jun}-1999 10:42 &{\bf interacting}&-28 & 4.8 &450& 0.55 & 1.60 & 0.21 & 0.09 & 58 & 0.95 &  \\
45&17-{\tt Jan}-2000 12:09 &{\bf perturbed}&-44 & 4.1 &398& 0.38 & 2.01 & 1.05 & 0.51 & 68 &-0.08 &  \\
46&27-{\tt Nov}-2001 19:47 &{\bf interacting}& 73 & 2.3 &779& 0.25 & 2.34 & 0.23 & 0.16 & 37 & 0.39 &  \\
 \hline

\end{tabular}
\end{center}

\begin{list}{}{}
\item[$^{\mathrm{a}}$] Number identifying MCs.
\item[$^{\mathrm{b}}$] $t_c$ is the time of closest approach from the MC axis.
\item[$^{\mathrm{c}}$] The MCs are separated in three groups: non-perturbed,
                        perturbed, and in interaction ({denoted interacting}).
\item[$^{\mathrm{d}}$] $\theta$ and $D$ are the latitude and solar distance.
\item[$^{\mathrm{e}}$] $V_c$ is the velocity at the closest distance from the MC axis and $S$
                        is the MC size. Both are computed in the radial direction away from the Sun ($\uvec{R}$).
\item[$^{\mathrm{f}}$] $<B>$, $<\Np>$ and $<\beta_p>$ are the average over the flux rope of
                         the field strength, the proton density and the proton $\beta$,
                         respectively.
\item[$^{\mathrm{g}}$] $\gamma$ is the acute angle between the MC axis and the radial
                         direction ($\uvec{R}$).
\item[$^{\mathrm{h}}$] $\zeta$ is the unidimensional expansion rate [Eq.~(\ref{zeta})]. $\zeta<0$
  means a MC observed in compression stage.
\end{list}

\end{table*}
%%%%%%%%%%%%%%%%%%% TABLE %%%%%%%%%%%%%%%%%%% TABLE %%%%%%%%%%%%%%%%%%%%%%%%

\end{appendix}

\begin{acknowledgements}

% France/Argentina
The authors acknowledge financial support from ECOS-Sud
through their cooperative science program (N$^o$ A08U01).
%Grants
This work was partially supported by the Argentinean grants:
UBACyT 20020090100264, PIP 11220090100825/10 (CONICET), and PICT-2007-856
(ANPCyT).
% Conicet
S.D. is member of the Carrera del Investigador Cien\-t\'\i fi\-co, CONICET.
% ICTP
S.D. acknowledges support from the Abdus Salam International Centre
for Theoretical Physics (ICTP), as provided in the frame of his
regular associateship.
% Luciano
{L.R. acknowledges support from the Belgian Federal Science Policy
Office through the ESA-PRODEX program, and the European Union
Seventh Framework Programme (FP7/2007-2013) under grant agreement
number 263252 [COMESEP].}

\end{acknowledgements}

%%%%%%%%%%%%%%%%%%%%%%%%%%%%%%%%%%%%%%%%%%%%%%%%%%%%%%%%%%%%%%%%%%%%%
% format of references provided by the review (.bst)
\bibliographystyle{aa}
      % file containing the bibtex references (.bib)
\bibliography{mc}
      % look if the file containing the ``\bibitem'' exits
\IfFileExists{\jobname.bbl}{}
{\typeout{}
\typeout{****************************************************}
\typeout{****************************************************}
\typeout{** Please run "bibtex \jobname" to optain}
\typeout{** the bibliography and then re-run LaTeX}
\typeout{** twice to fix the references!}
\typeout{****************************************************}
\typeout{****************************************************}
\typeout{}
}

\end{document}